\documentclass[a4paper,superscriptaddress,eqsecnum,aps,prd,nofootinbib, notitlepage, twocolumn, showpacs]{revtex4-1}

\usepackage{fancyhdr}
\usepackage[a4paper,left=15mm,right=15mm, top=2cm, bottom=2cm]{geometry}
\usepackage{amssymb, amsmath}
\usepackage{graphicx}
\usepackage{verbatim}
\usepackage{booktabs}
\usepackage{placeins}
\usepackage{float}
\usepackage{pdfpages}

\usepackage{tocloft}
\usepackage[T1]{fontenc}
\newcommand{\changefont}[3]{\fontfamily{#1} \fontseries{#2} \fontshape{#3} \selectfont}

\usepackage[plainpages=false, pdfpagelabels]{hyperref}
\hypersetup{colorlinks=true, linkcolor=black, citecolor=black, urlcolor=black}

\changefont{ptm}{m}{n}

\graphicspath{{figures/}}

\begin{document}

\title[Relativistic virialization in the Spherical Collapse model]{\textbf{\Large{Relativistic virialization in the Spherical Collapse model for Einstein-de Sitter and \texorpdfstring{$\mathbf{\Lambda}$}{[Lambda]}CDM cosmologies}}}

\author{Sven Meyer}
\email{sven.meyer@stud.uni-heidelberg.de}
\affiliation{Institut f\"ur theoretische Astrophysik (ITA), Zentrum f\"ur Astronomie, Universit\"at Heidelberg, Albert-Ueberle-Stra\ss e 2, 69120 Heidelberg, Germany}

\author{Francesco Pace}

\affiliation{Institute of Cosmology and Gravitation (ICG), University of Portsmouth, Dennis Sciama Building, Burnaby Road, Portsmouth PO1 3FX, United Kingdom}

\author{Matthias Bartelmann}

\affiliation{Institut f\"ur theoretische Astrophysik (ITA), Zentrum f\"ur Astronomie, Universit\"at Heidelberg, Albert-Ueberle-Stra\ss e 2, 69120 Heidelberg, Germany}


\date{\footnotesize{Received \today; accepted ?}}

\pagestyle{fancy}

\begin{abstract}
\noindent  Spherical collapse has turned out to be a successful semi-analytic model to study structure formation in different DE models and theories of gravity, but nevertheless, the process of virialization is commonly studied on the basis of the virial theorem of classical mechanics. In the present paper, a fully generally-relativistic virial theorem based on the Tolman-Oppenheimer-Volkoff (TOV) solution for homogeneous, perfect-fluid spheres is constructed for the Einstein-de Sitter and $\Lambda$CDM cosmologies. We investigate the accuracy of classical virialization studies on cosmological scales and consider virialization from a more fundamental point of view. Throughout, we remain within general relativity and the class of FLRW models. The virialization equation is set up and solved numerically for the virial radius, $y_\mathrm{vir}$, from which the virial overdensity $\Delta_V$ is directly obtained. Leading order corrections in the post-Newtonian framework are derived and quantified. In addition, problems in the application of this formalism to dynamical DE models are pointed out and discussed explicitly. We show that, in the weak field limit, the relative contribution of the leading order terms of the post-Newtonian expansion are of the order of $10^{-3}\%$ and the solution of Wang \& Steinhardt (1998) (see \cite{wang_cluster_1998}) is precisely reproduced. Apart from the small corrections, the method could provide insight into the process of virialization from a more fundamental point of view.
\end{abstract}

\maketitle

\small

\section{Introduction}\label{int}

The question of how structures form in the Universe is a long-standing topic in theoretical cosmology and provides a lot of room for discussion. Since the fully non-linear regime cannot be accessed analytically, huge N-body simulations have been set up to describe structure formation by gravitationally interacting particles in an expanding background. However, these attempts are computationally costly and therefore perturbative approaches have been developed in order to keep the continuous character of general relativity and the FLRW model and make use of methods from fluid mechanics. A very simple semi-analytic model of this kind is the \textbf{Spherical Collapse}. A spherical, overdense patch evolves with the background expanding universe, slows down due to its self-gravity, turns-around and collapses. The object is stabilized by virialization which prevents it from collapsing into a singularity. Despite its simplicity and idealizations, this model gives a first insight into the formation of spherical halos at all mass scales. The underlying formalism dates back to Gunn and Gott in 1972 (see \cite{gunn_infall_1972}), but has been rediscovered and continuously extended in recent years (see \cite{abramo_structure_2007, bartelmann_non-linear_2006, basilakos_virialization_2007, lee_spherical_2010, manera_cluster_2006, maor_spherical_2007, maor_virialization_2005, mota_spherical_2004, nunes_structure_2006, pace_spherical_2010, wang_cluster_1998, wang_virialization_2006, wintergerst_clarifying_2010}).\\

In this work, we are going to use the results of Pace et al. (2010) (see \cite{pace_spherical_2010}) to investigate the process of virialization and try to find answers to some remaining questions in this context. The virial theorem provides a powerful tool to study systems in equilibrium, but in order to clarify its role in the framework of general relativity, a relativistic version is needed. After giving an overview of the classical concepts and the requirements of relativistic calculations in Sects. \ref{sec:class} and \ref{sec:Req}, we will derive a relativistic version of the virial theorem based on the Tolman-Oppenheimer-Volkhoff equation (see \cite{oppenheimer_massive_1939, tolman_static_1939} for the original references) in an Einstein-de Sitter and $\Lambda$CDM universe (see Sects. \ref{sec:TOV_L},  \ref{sec:vir_rel}). In the following, this will be applied to the virialization equation in the spherical collapse model and a post-Newtonian expansion will be performed (see Sects. \ref{sec:vir_eq}, \ref{sec:PN}). The relativistically corrected results for the virial radius and virial overdensity will be discussed and leading order corrections are worked out in particular. (Sect. \ref{sec:res}). We will also dedicate a section to the problems occurring when this formalism is applied to general DE models and point out possible ideas to solve them (see Sect. \ref{sec:dyn_DE}). Throughout the paper, we will make use of natural units, i. e. $c=1$.

\section{Virialization in the classical spherical collapse}\label{sec:class}

In the classical treatment of virialization, there are two major ingredients that have to be well-understood. First of all, structure formation in the present universe is highly non-linear on scales less than $10$ Mpc and an evolution equation for the spherical patch is needed that takes this non-linearity into account. Secondly, the virial theorem has to be combined with energy conservation to a virialization condition that allows determining the time when collapse stops and the system reaches an equilibrium. The key quantities, that are assigned to it, are the virial radius normalized to the turn-around one, $y_{\mathrm{vir}}= R_{\mathrm{vir}}/R_{\mathrm{ta}}$, and the virial overdensity with respect to the background, $\Delta_{V} = \rho(R_{\mathrm{vir}})/\rho_{\mathrm{b}}(a_{\mathrm{vir}})$. These are general functions of redshift and provide a characterization of the equilibrium state of the halo. \\

The non-linear evolution equation of a spherical overdensity of pressureless dark matter has already been treated in detail by many authors (see for example \cite{abramo_structure_2007, ohta_evolution_2003, pace_spherical_2010}). The resulting equation

\begin{equation}
\label{nlm}
 \delta'' +\left(\frac{3}{a}+\frac{E'}{E}\right)\delta'-\frac{4}{3}\frac{\delta'^2}{1+\delta}-\frac{3}{2}\frac{\Omega_\mathrm{m,0}}{a^5E^2(a)}\delta \left(1+\delta\right) = 0 
\end{equation}

describes the non-linear evolution of a spherical, top-hat density contrast $\delta(a)=\frac{\rho-\rho_b}{\rho_b}$ with respect to the background dark matter density $\rho_b$. $E(a)$ contains all the dynamics of the background cosmological model and is related to the Hubble function via the expression $H(a)= H_0 E(a)$. It will be important to express the virial overdensity $\Delta_V$ as a function of the turn-around one denoted by $\zeta$: 

\begin{equation}
\label{Delta_V}
\begin{split}
&\Delta_V = 1 + \delta(a_\mathrm{vir}) = \zeta \left(\frac{x_\mathrm{vir}}{y_\mathrm{vir}}\right)^3 \\ 
&\text{with} \quad x_\mathrm{vir}=\frac{a_\mathrm{vir}}{a_\mathrm{ta}}, \ y_\mathrm{vir}=\frac{R_\mathrm{vir}}{R_\mathrm{ta}}.
\end{split}
\end{equation}

The virial radius, $y_\mathrm{vir}$, is obtained from the virialization equation in which the classical virial theorem $\overline{T}= \overline{\dfrac{R}{2}\dfrac{\partial U}{\partial R}}$ is combined with the assumption of energy conservation during collapse.\footnote{In the following, we will drop the bars over the time averaged quantities and implicitly assume time averaging.} It should be mentioned that energy conservation is a very common assumption in the literature and it is not proven whether it can actually be applied. Maor \& Lahav (2005), as well as Wang (2006), pointed out that a homogeneous DE component with $w\neq -1, \ -1/3$ clearly violates energy conservation between turn-around and collapse (see \cite{maor_spherical_2007, maor_virialization_2005, wang_virialization_2006})

 \begin{equation}
\label{vir_eq}
 \left[U+\frac{\partial U}{\partial R}\right]_{\mathrm{vir}}=U_{ \mathrm{ta}}.
\end{equation}

In case of an Einstein-de Sitter universe, one simply obtains $y_\mathrm{vir}= \frac{1}{2}$ whereas the corresponding virial overdensity (evaluated at collapse scale factor) is given by $\Delta_V = 18 \pi^2 \sim 178$ (see \cite{wang_virialization_2006}). In case of $\Lambda$CDM and dynamical DE models, two major classical results have been proposed in the literature:

\begin{setlength}{\leftmargini}{0.3cm}
\begin{itemize}
 \item Wang \& Steinhardt (1998) (WS afterwards)  (see \cite{wang_cluster_1998})\footnote{Horellou \& Berge (2005) (\cite{horellou_dark_2005}) have proposed a generalization due to dynamical DE models, but in the $\Lambda$CDM model both results agree.} :
\begin{equation} 
\label{WS_result}
 y_{\mathrm{vir}}=\frac{1-\frac{\eta_v}{2}}{2+\eta_t-\frac{3}{2}\eta_v}
\end{equation}
\item Wang (2006) (PW afterwards) (see \cite{wang_virialization_2006}):
\begin{equation}
\label{PW}
y_{\mathrm{vir}}= \frac{1-(1+3w)\frac{\eta_t}{2}}{2-(1+3w)\frac{\eta_t}{2}} \underset{(w=-1)} {=} \frac{1+\eta_t}{2+\eta_t}
\end{equation}
\end{itemize}
\end{setlength}

with the WS parameters $\eta_v$ and $\eta_t$, and the equation-of-state-parameter $w$ given by

\begin{eqnarray*}
\eta_t&=&2\zeta^{-1}\frac{\Omega_{\mathrm{\Lambda}}(a_{ \mathrm{ta}})}{\Omega_{\mathrm{m}}(a_{ \mathrm{ta}})} \\ 
\eta_v&=&2\zeta^{-1}\left(\frac{a_{ \mathrm{ta}}}{a_{\mathrm{vir}}}\right)^3\frac{\Omega_{\mathrm{\Lambda}}(a_{\mathrm{vir}})}{\Omega_{\mathrm{m}}(a_{\mathrm{vir}})}\\
w &=& \frac{p_{\Lambda}}{\rho_{\Lambda}}=-1.
\end{eqnarray*}

The corresponding virial overdensities become functions of the collapse (virial) scale factor $a_c$ ($a_\mathrm{vir}$) and reach the EdS value asymptotically for small scale factors ($a_c < 10^{-1}$) corresponding to the matter dominated era.\footnote{It has to be mentioned for completeness that this is only true for dynamical DE models that have negligible contribution in the matter dominated era. Counterexamples are early DE models (see results in \cite{pace_spherical_2010} and references therein).}

\section{Requirements for relativistic calculations}\label{sec:Req}

Relativistic treatment of virialization in the same way as done in the classical case causes some trouble, because energy conservation is not global in general relativity. A second problem has been addressed by Komar (see \cite{komar_asymptotic_1962, komar_positive-definite_1963}) stating that isolated bodies like a spherical halo can only be described exactly in asymptotically flat spacetimes which is generally not given in case of FLRW models. A promising way out of these problems is assuming that the scale of the halo is much smaller than the typical length scale of the background universe given by the Hubble radius. If the Killing vector field of the FLRW spacetime is considered with respect to this assumption, its time-like component greatly exceeds its spatial components, allowing to neglect the latter and at good approximation consider the Killing vector as time-like.

\begin{figure*}
\centering
 \includegraphics[width=15cm]{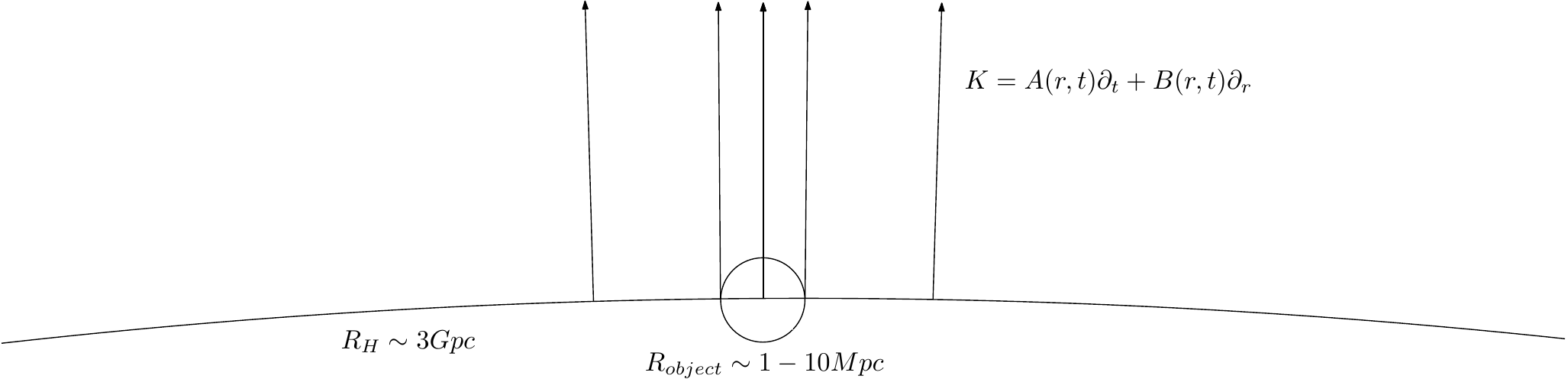}
 \caption[Killing vector field $K$, small scales]{Killing vector field on the spacelike hypersurface of the universe compared to an object of much smaller radius}
  \label{radii}
\end{figure*}

Fig. (\ref{radii}) illustrates that neighbouring Killing vectors are approximately parallel on the scale of the object, which means that the radial component of $K$ is extremely small on these scales and thus negligible. A detailed quantitative analysis of this issue is given in Appendix \ref{app_A}.\\
\\
From this argument, we can infer three major conclusions essential for the following considerations:   

\begin{setlength}{\leftmargini}{0.3cm}

\begin{itemize}
 \item 
  Since $r\ll R_H$, the solutions can be considered as nearly asymptotically flat such that isolated objects can be defined in GR and the halo mass $M$ is well-defined in the sense of a Komar integral. 
  \item
  Approximately, one can define the scale of the object as local and introduce a local coordinate frame which allows energy-momentum conservation, $\nabla_\nu T^{\mu\nu}=0$, to hold in that frame. Since the virialization equation, $T_\mathrm{vir}+U_\mathrm{vir}=U_\mathrm{ta}$, represents energy conservation, this condition is essential. 
\item 
  The approximately time-like Killing vector field $K$ of Eq. (\ref{metric}) allows static solutions of the field equations within an accuracy of $\sim 0.3 \ \%$ (see Eq. (\ref{AB})). In addition, the typical structure of the FLRW metric, especially the fact that a global time coordinate can be introduced and a space-like foliation with it, infers the orthogonality of $K$ to the underlying hypersurfaces such that the Frobenius condition\footnote{For $\omega$ being the corresponding dual vector to $K$, the Frobenius condition states that $\omega \wedge d\omega = 0$ which turns out to be equivalent to $K$ being orthogonal to the spacelike sections spanned by suitably chosen spatial coordinates. (see \cite{straumann_general_2004})} is naturally fulfilled. Thus, we can insert and compare static solutions at turn-around and virial redshift in the virialization equation $T_\mathrm{vir}+U_\mathrm{vir}=U_\mathrm{ta}$. Since the virial theorem will be applied to the final state which is static by definition (within the timescales we consider), this approximation holds in particular for the turn-around being a critical point, but not a static one in the exact treatment. 
\end{itemize}
\end{setlength}

\section{TOV equation for matter in the presence of a cosmological constant}\label{sec:TOV_L}

Relying on the assumptions of the previous section, we can set up a spherically symmetric, static spacetime with the metric:

\begin{equation}
\label{metric2}
 ds^2=-e^{2a(r)}\mathrm{d}t^2+e^{2b(r)}\mathrm{d}r^2+r^2d\Omega^2.
\end{equation}

In addition, we consider a system of two fluid components described by

\begin{equation}
  T_{\mu\nu}= T_{\mu\nu}^{(m)}+ T_{\mu\nu}^{(\Lambda)} = (\rho_T+p_T)u_\mu u_\nu +p_Tg_{\mu\nu}
\end{equation}

in which

\begin{equation*} 
 \rho_T=\rho_m+\rho_{\Lambda}, \quad  p_T=p_m+p_{\Lambda}.
\end{equation*}

The equation of state of the cosmological constant fluid corresponds to $w=\frac{p_{\Lambda}}{\rho_{\Lambda}}=-1$.
 
Energy and momentum are locally conserved for the total system as well as for each component separately which means 

\begin{equation*}
 \nabla_\nu T^{\mu\nu}=0   \quad \textnormal{and} \quad \nabla_\nu T^{\mu\nu (m)}=0 , \quad \nabla_\nu T^{\mu\nu (\Lambda)}=0. 
\end{equation*}

 Projecting the conservation equation of the (clustering) matter component onto the space perpendicular to the time direction leads to the relativistic Euler equation for the matter fluid

\begin{equation}
 h_{\alpha\mu}\nabla_\nu T^{\mu\nu (m)}=0 \quad \textnormal{with} \quad h_{\alpha\mu}=g_{\alpha\mu}+u_\alpha u_\mu. 
\end{equation}
 
Working that out, one finds \footnote{In the following sections, we define $\rho_m\equiv\rho
$ and $p_m \equiv p$.}

\begin{equation}
\label{Euler_rel}
 (\rho+p)\nabla_u u=-\nabla_\alpha p + u \nabla_u p.
\end{equation}

In case of a static configuration ($\nabla_u p=0$) and with the help of Eq. (\ref{metric2}), Eq. (\ref{Euler_rel}) reduces to:

\begin{equation}
\label{equi}
 a'=\frac{-p'}{\rho+p}.
\end{equation}

The field equations for the given metric are

\begin{eqnarray} \nonumber
 G_{\mu\nu}=8\pi G T_{\mu\nu}\\ \nonumber \\
\label{feqs 1}
\frac{1}{r^2}-e^{-2b}\left(\frac{1}{r^2}-\frac{2b'}{r}\right) &=& 8\pi G (\rho+\rho_{\Lambda}) \\
\label{feqs 2}
-\frac{1}{r^2}+e^{-2b}\left(\frac{1}{r^2}+\frac{2a'}{r}\right) &=& 8\pi G (p-\rho_{\Lambda}) \\
\label{feqs 3}
e^{-2b}\left(a''-a'b'+a'^2+\frac{a'-b'}{r}\right) &=& 8\pi G (p-\rho_{\Lambda}).
\end{eqnarray}

Combining Eqs. (\ref{feqs 1}) and (\ref{feqs 2}) with (\ref{equi}) leads to 

\begin{eqnarray}
 e^{-b}&=&\frac{1}{\sqrt{1-\frac{8\pi G}{3}(\rho+\rho_{\Lambda})r^2}}\\
\label{TOV_rel}
-p'&=&\frac{4\pi G}{3} r \cdot \frac{(\rho+p)(\rho+3p-2\rho_{\Lambda})}{1-\frac{8\pi G}{3}(\rho+\rho_{\Lambda})r^2}.
\end{eqnarray}

Eq. (\ref{TOV_rel}) is the TOV equation for the $\Lambda$CDM-model in case of homogeneous densities. For solving it, we assume the matter pressure to vanish at the boundary $p(r=R)=0$.\footnote{In the derivation of Eq. (\ref{nlm}), the assumption of pressureless dark matter is a crucial argument. Nevertheless, for consistency with the TOV equation, we have to allow a pressure profile for the interior of the sphere. This issue will be discussed below.}\\

This leads to

\begin{equation}
\label{pressure}
 p(r)=\frac{\rho\left(f \left(\frac{1-Ar^2}{1-AR^2}\right)^{1/2}-1\right)+2 \rho_{\Lambda}}{3-f \left(\frac{1-Ar^2}{1-AR^2}\right)^{1/2}}
\end{equation}

where

\begin{equation}
 A=\frac{8 \pi G}{3} \left(\rho+\rho_{\Lambda}\right), \quad  f=1- \frac{2\rho_{\Lambda}}{\rho}, \quad  \rho_{\Lambda} = \frac{\Lambda}{8 \pi G}. 
 \end{equation}

Inserting Eq. (\ref{TOV_rel}) and Eq. (\ref{pressure}) into the hydrostatic equilibrium condition (Eq. (\ref{equi})) and integrating with the boundary $a(R)=1/2\ln(1-AR^2)$ (Schwarzschild-de Sitter solution) gives

\begin{equation}
 e^{a}=(1-AR^2)^{1/2}\frac{3-\left(\frac{1-Ar^2}{1-AR^2}\right)^{1/2}f}{3-f}
\end{equation}

and thus the full metric inside the sphere can be written as

\begin{equation}
\label{metric_int}
\begin{split}
  ds^2=&-(1-AR^2)\left(\frac{3-\left(\frac{1-Ar^2}{1-AR^2}\right)^{1/2}f}{3-f}\right)^2 \mathrm{d}t^2\\
       &+\frac{\mathrm{d}r^2}{1-Ar^2}+r^2d\Omega^2
\end{split}
\end{equation}

which represents the metric of the interior Schwarzschild-de Sitter spacetime. \\

The well-known exterior Schwarzschild-de Sitter solution 

\begin{equation}
\begin{split}
ds^2 = &-\left(1-\frac{2GM}{r} -\frac{\Lambda r^2}{3} \right) \mathrm{d}t^2 + \frac{\mathrm{d}r^2}{1-\frac{2GM}{r} -\frac{\Lambda r^2}{3}} \\
       &+ r^2 d\Omega^2
\end{split}
\end{equation}

matches continuously with Eq. (\ref{metric_int}) at $r=R$. In this particular case, it has to be mentioned that asymptotic flatness can only be reached approximately as discussed in Sect. (\ref{sec:Req}). Since the scale of the halo is much smaller than the Hubble radius ($r/R_H \sim 10^{-3}$) we can still assume the object to be nearly isolated. We decided to embed the sphere into the Schwarzschild-de Sitter spacetime instead of an FLRW spacetime, because spacetime around the object can be assumed to be approximately static as well (due to the approximated time-like Killing vector field on these scales). In the ordinary Tolman-Oppenheimer-Volkhoff solution (see \cite{oppenheimer_massive_1939}), the perfect fluid sphere is embedded into vacuum described by the Schwarzschild solution. In order to be consistent with this approach, the generalization including a cosmological constant is embedded into the Schwarzschild-de Sitter spacetime. Nevertheless, it will turn out that the virial radius and overdensity can be predicted consistently with this approach, although a dark matter contribution outside the sphere is neglected (see Sect. (\ref{sec:res}) and the weak field limits in Sects. (\ref{sec:vir_eq}) and (\ref{sec:PN})).     

\section{Derivation of the relativistic virial theorem}\label{sec:vir_rel}

The pressure profile in Eq. (\ref{pressure}) contains the radial dependence of the pressure in a sphere consisting of a cosmological constant fluid and collapsed dark matter embedded into a background Schwarzschild-de Sitter spacetime. When virialization starts, the system can be approximately assumed to be in equilibrium which means that it can really be described by Eq. (\ref{TOV_rel})\footnote{The TOV-equation represents the equation of motion of the system in equilibrium.}. In order to derive a virial theorem from that, one can take the first spatial moment which should usually lead to the virial theorem after time averaging. This means that small fluctuations around the equilibrium state are averaged out over time such that only time averaged quantities (energy expressions) are left in the virial theorem. Since the system is already in equilibrium and the TOV equation has no time-dependences, the time integral drops naturally and all quantities can be interpreted as time-averaged.\\
 
Eq. (\ref{TOV_rel}) is multiplied with $r$ and integrated (averaged) over the spacetime volume element (hence taking the spatial moment and time-averaging are performed in one step):

\begin{equation}
\begin{split}
 -&\lim_{T\rightarrow \infty}\frac{1}{T}{\int{p' r\eta}}\\
=&\lim_{T\rightarrow \infty}\frac{1}{T}{\int {\eta \left[\frac{4\pi G}{3} r^2 \frac{(\rho+p)(\rho+3p-2 \rho_{\Lambda})}{1-Ar^2}\right]}}
\end{split}
\end{equation}

which becomes

\begin{equation}
 \begin{split}
   -&4 \pi\lim_{T\rightarrow \infty}\frac{1}{T} \int_{0}^{T}\int_{0}^{R}{p' r^3 e^{a+b} \mathrm{d}t\mathrm{d}r} \\
  =\ &4\pi \lim_{T\rightarrow \infty}\frac{1}{T} \int_{0}^{T}\int_{0}^{R} { \left[\frac{4\pi G}{3} r^4 \cdot \frac{(\rho+p)(\rho+3p-2\rho_{\Lambda})}{1-Ar^2} \right.}\\
   & \left. e^{a+b}\right] \mathrm{d}t \mathrm{d}r.
 \end{split}
\end{equation}

Since all the quantities in the integral do not depend on time, the evaluation of the time integral cancels naturally and, while interpreting the given quantities as time-averaged, this becomes

\begin{equation}
 \begin{split}
   &-4 \pi\int_{0}^{R}{p' r^3 e^{a+b} \mathrm{d}r} \\
  &= 4\pi \int_{0}^{R} { \left[\frac{4\pi G}{3} r^4 \cdot \frac{(\rho+p)(\rho+3p-2\rho_{\Lambda})}{1-Ar^2}e^{a+b}\right] \mathrm{d}r}.
 \end{split}
\end{equation}

Looking at the LHS of this equation in Euclidean space and performing a partial integration, we see that:

\begin{eqnarray}
   -4 \pi\int_{0}^{R}{p' r^3  \mathrm{d}r}&=&\left[-4 \pi p r^3\right]_\mathrm{0}^R+\int_\mathrm{0}^R{12\pi r^2 \mathrm{d}r} \\
 &=& \int{3p \mathrm{d}V}\\
 &\equiv&2 \overline{T}.
\end{eqnarray}

In consistency with the Euclidian case, we can propose that 

\begin{equation}
\label{pres}
  2T \equiv 2\overline{T}= -4 \pi\int_{0}^{R}{p' r^3 e^{a+b} \mathrm{d}r}.
\end{equation}

Consequently, we define:

\begin{equation}
\label{2T}
   2T  =  4\pi \int_{0}^{R} { \left[\frac{4\pi G}{3} r^4 \cdot \frac{(\rho+p)(\rho+3p-2\rho_{\Lambda})}{1-Ar^2}e^{a+b}\right] \mathrm{d}r}.
\end{equation}

Inserting Eq. (\ref{pressure}) and Eq. (\ref{metric2}) into Eq. (\ref{2T}), we obtain

\begin{equation}
\begin{split}
  2T \ = \ &\frac{16 \pi^2 G}{3(3-f)}\left(1-AR^2\right)^{1/2}\left(2\rho+2\rho_{\Lambda}\right)^2\\ &\int_\mathrm{0}^R{\frac{r^4}{(1-Ar^2)^{3/2}}\cdot \frac{\left(\frac{1-Ar^2}{1-AR^2}\right)^{1/2}f}{3-\left(\frac{1-Ar^2}{1-AR^2}\right)^{1/2}f}\mathrm{d}r}.
\end{split}
\end{equation}

This is one version of a fully relativistic virial theorem for clustering dark matter in a $\Lambda$CDM-background model. Of course, other attempts exist in the literature to derive a relativistic virial theorem for several purposes. Chandrasekhar (\cite{chandrasekhar_post-newtonian_1965}) derived a post-Newtonian version of the tensor virial theorem by investigating the post-Newtonian hydrodynamic equations consistently with Einstein's field equations. Bonazzola (1973) (\cite{bonazzola_virial_1973}) has proposed an integral identity consistent with general relativity in an asymptotically flat, stationary and axisymmetric spacetime. Vilain (1979) \cite{vilain_virial_1979} considers a scalar generalization of the virial theorem to general relativity which is valid for spherically symmetric, asymptotically flat spacetimes which has been successfully applied to stability studies of perfect fluid spheres. In addition, Vilain's work allows to interprete the result of Bonnazzola (1973)  geometrically in the spherical case. Gourgoulhon \& Bonazzola (1994) (\cite{bonazzola_virial_1994}) extended the work of 1973 to any stationary, asymptotically flat spacetime in general. Straumann (\cite{straumann_general_2004}) proposes a virial expression in case of a spherically symmetric, static spacetime based on the Komar integral and asymptotic flatness. Except (\cite{chandrasekhar_post-newtonian_1965}), these remarkable results have in common that asymptotical flatness is a crucial assumption to the spacetime which is necessary in order to define isolated objects in the sense of a Komar integral (see \cite{komar_asymptotic_1962, komar_positive-definite_1963}). We want to emphasize at this point that, strictly speaking, this condition has to be valid in our case as well. However, we make use of the fact that an isolated object can be approximately defined in the FLRW spacetime by assuming the scale of the halo to be much smaller than the corresponding Hubble radius.

\section{Relativistic gravitational potential energy}\label{sec:pot_rel}

The modified TOV solution can also be applied to find a relativistic expression for the gravitational potential energy of a spherical body. The derivation is inspired by the considerations of N. Straumann  (see \cite{straumann_general_2004}), but since it is quite technical, we refer to Appendix \ref{app_B} and quote here only the final result:

\begin{eqnarray}
\label{T}
  T &=& \int_\mathrm{0}^R {4 \pi r^2 \epsilon \frac{1}{\sqrt{1-Ar^2}}\mathrm{d}r} \\
\label{U}
  U &=& \int_\mathrm{0}^R {4 \pi r^2 \rho \left(1- \frac{1}{\sqrt{1-Ar^2}} \right) \mathrm{d}r}
\end{eqnarray}

with $\epsilon$ denoting the relativistic kinetic energy density which is defined in Eq. (\ref{epsilon}) (see Appendix \ref{app_B}).

In case of small gravitational fields given for an object having a radius $r$ which is much larger than its Schwarzschild radius (this corresponds to $Ar^2 \ll 1$), we can expand Eq. (\ref{T}) and Eq. (\ref{U}) to first order:

\begin{eqnarray*} 
  T &=& \frac{4 \pi R^3}{3} \epsilon + \frac{2\pi \epsilon A R^5}{5} + \mathcal{O}(A^2)\\ 
    &=& M- M_\mathrm{0} + \frac{2\pi \epsilon A R^5}{5} + \mathcal{O}(A^2)\\
  U &=& -\frac{2 \pi \rho A}{5} R^5 + \mathcal{O}(A^2) \\ 
    &=& -\frac{3}{5} \frac{G M^2}{R} -\frac{4 \pi G}{5} M \rho_{\Lambda} R^2 + \mathcal{O}(A^2).
\end{eqnarray*}

The kinetic energy will reduce to the specially-relativistic result if gravitational effects are neglected to zeroth order. The potential energy contains the Newtonian self-energy of a homogeneous sphere as a leading-order term. Thus, classical limits can be reproduced showing that Eq. (\ref{T}) and Eq. (\ref{U}) are consistently defined.

\section{Virialization equation} \label{sec:vir_eq}

Assuming that energy conservation still holds during collapse, the virialization equation states

\begin{equation}
\label{virialization_eq}
 \left[T+U\right]_\mathrm{vir}=U_\mathrm{ta}.
\end{equation}

Let us now insert all derived energy expressions and perform a change of variable $r\rightarrow y \equiv r/R_\mathrm{ta}$. After simplifying the result, we end up with

\begin{equation}
\label{vir}
 \begin{split}
  T_\mathrm{vir}=&\frac{8\pi^2 G R_\mathrm{ta}^5 }{3(3-f)}\left(1-A_\mathrm{vir}y_\mathrm{vir}^2R_\mathrm{ta}^2\right)^{1/2}\left(2\rho_\mathrm{vir}+2\rho_{\Lambda}\right)^2  \\ &\int_\mathrm{0}^{y_\mathrm{vir}}{\frac{y^4}{\left(1-A_\mathrm{vir}y^2R_\mathrm{ta}^2\right)^{3/2}} \cdot \frac{\left(\frac{1-A_\mathrm{vir}y^2R_\mathrm{ta}^2}{1-A_\mathrm{vir}y_\mathrm{vir}^2R_\mathrm{ta}^2}\right)^{1/2}f}{3-\left(\frac{1-A_\mathrm{vir}y^2R_\mathrm{ta}^2}{1-A_\mathrm{vir}y_\mathrm{vir}^2R_\mathrm{ta}^2}\right)^{1/2}f}dy}
  \end{split}
\end{equation}

\begin{equation}
\label{pot_vir}
U_\mathrm{vir}=4\pi \rho_\mathrm{vir} R_\mathrm{ta}^3\int_{0}^{y_\mathrm{vir}}{y^2\left(1-\frac{1}{\sqrt{1-A_\mathrm{vir}y^2R_\mathrm{ta}^2}}\right)dy} 
\end{equation}

\begin{equation}
\label{pot_ta}
U_\mathrm{ta}=4\pi \rho_\mathrm{ta} R_\mathrm{ta}^3\int_{0}^{1}{y^2\left(1-\frac{1}{\sqrt{1-A_\mathrm{ta}y^2R_\mathrm{ta}^2}}\right)dy} 
\end{equation}

with the definitions 

 \begin{equation*} 
\begin{split}
 &A_\mathrm{vir}=\frac{8 \pi G}{3} \left(\rho_\mathrm{vir}+\rho_{\Lambda}\right), \quad  A_\mathrm{ta}=\frac{8 \pi G}{3} \left(\rho_\mathrm{ta}+\rho_{\Lambda}\right), \\
 &f=1 - \frac{2\rho_{\Lambda}}{\rho_\mathrm{vir}}, \quad \rho_\mathrm{vir} = \frac{3M}{4 \pi y_\mathrm{vir}^3 R_\mathrm{ta}^3}, \quad \rho_\mathrm{ta} = \frac{3M}{4 \pi R_\mathrm{ta}^3}. 
\end{split}
\end{equation*}

Eq. (\ref{virialization_eq}) has to be solved numerically for $y_\mathrm{vir}$ at different redshifts (see Fig. \ref{relat_vir} for the results). The turn-around radius, $R_\mathrm{ta}$, can be obtained by using the solution of Eq. (\ref{nlm}) and Eq. (\ref{Delta_V}):

\begin{equation}
\label{R_ta}
\begin{split}
 &\zeta =  \frac{\rho(R_\mathrm{ta})}{\rho^b_m(a_\mathrm{ta})} = 1+\delta (a_\mathrm{ta}), \quad \rho(R_\mathrm{ta})=\frac{3 M}{4\pi R_\mathrm{ta}^3}, \\
&\rho^b_m(a_\mathrm{ta}) = \frac{3H_0^2
}{8 \pi G} \Omega^{(0)}_m a_\mathrm{ta}^3 \\
&\Rightarrow \ R_\mathrm{ta}=a_\mathrm{ta}\cdot \left( \frac{2GM}{H_{0}^2\Omega_m^{(0)}(1+\delta(a_\mathrm{ta}))}\right)^{1/3}.
\end{split}
\end{equation}\\
 
Let us consider the classical limit with respect to two assumptions:

\begin{setlength}{\leftmargini}{0.3cm}
\begin{itemize}
 \item 
  The sphere's radius is much larger than its Schwarzschild radius $R_S= AR^3 \ll  R$, i.e. $AR^2 \ll 1$.
\item
  The cosmological-constant density is much smaller than the dark matter density inside the sphere. Since $\rho_{\Lambda}$ is of the order of the critical density and $\rho=\Delta_V \rho_{cr}$ with $\Delta_V\sim 95 - 180$\footnote{See Pace et al. (2010) (\cite{pace_spherical_2010}) for their results in the $\Lambda$CDM case.},  this can be assumed safely in our case. 
\end{itemize}
\end{setlength}

Expanding Eq. (\ref{vir}) to first order in $AR^2$ and $\rho_{\Lambda}/\rho$, and simplifying it, we end up with

\begin{equation}
 y_\mathrm{vir}^2=\frac{\rho_\mathrm{ta}+\rho_{\Lambda}}{\frac{1}{2}\rho_\mathrm{vir}+2 \rho_{\Lambda}}.
\end{equation}

Writing this in terms of the Wang-Steinhardt-parameters $\eta_t$ and $\eta_v$\footnote{See Wang \& Steinhardt (1998) (\cite{wang_cluster_1998}) and Sect. \ref{sec:class}}, this becomes

\begin{equation}
\label{WS}
 -2\eta_vy_\mathrm{vir}^3+\left(2+\eta_t\right)y_\mathrm{vir}-1=0.
\end{equation}

Eq. (\ref{WS}) can be solved approximately by

\begin{equation}
 y_\mathrm{vir}= \frac{1-\frac{\eta_v}{2}}{2+\eta_t-\frac{3}{2}\eta_v}.
\end{equation}

Thus, Eq. (\ref{vir}) reduces to the WS limit under the given assumptions.

\section{Post-Newtonian expansion of the virialization equation} \label{sec:PN}

The post-Newtonian expansion of the virialization equation can be done by simply performing a Taylor expansion of Eq. (\ref{vir}), however we choose a more elegant way including the equation of motion of the collapsing sphere.\footnote{This equation was first derived by Misner and Sharp (1964) (see \cite{misner_relativistic_1964}).}\\

We begin with a non-static, spherically symmetric spacetime described by

\begin{equation}
 ds^2=-e^{2a(r,t)}\mathrm{d}t^2+e^{2b(r,t)}\mathrm{d}r^2+r^2d\Omega^2
\end{equation}

and use again the energy-momentum tensor of an ideal fluid with two components

\begin{equation}
\begin{split}
 T_{\mu\nu}&=T_{\mu\nu}^{(m)}+T_{\mu\nu}^{(\Lambda)}\\
 &=(\rho_m+\rho_{\Lambda}+p_m+p_{\Lambda})u_{\mu}u_{\nu}+(p_m+p_{\Lambda})g_{\mu\nu}.
 \end{split}
\end{equation}

The $\Lambda$-component satisfies an equation of state given by

\begin{equation} 
\label{eos}
p_{\Lambda}=- \rho_{\Lambda}.
\end{equation}

Consider a comoving reference frame in which the four velocity $u$ has the components

\begin {eqnarray*} 
 u^0&=&e^{-a}\\ 
 u^i&=&0 \quad \textnormal {for} \ i=1,2,3.
\end {eqnarray*}

The relativistic Euler equation for the (clustering) matter component $h_{\alpha\mu}\nabla_{\nu}T^{\mu\nu, (m)}=0$ states (in that frame)\footnote{In the following, we define again $\rho_m \equiv \rho$, $p_m \equiv p$.}:

\begin{equation}
 a'=-\frac{p'}{\rho+p} \Rightarrow e^a=\frac{1}{\rho+p} \equiv \frac{1}{y}.
\end{equation}
 
If we combine this relation with the field equations for the metric, we obtain the relativistic equation of motion for a spherically symmetric object (first derived by Misner and Sharp (1964) in \cite{misner_relativistic_1964} and applied by Collins (1978) in \cite{collins_virial_1978})

\begin{equation}
\label{Eom}
\begin{split}
y\frac{d}{\mathrm{d}t}\left(y\frac{\mathrm{d}r}{\mathrm{d}t}\right)=&-\frac{1}{\rho+p}\frac{dp}{\mathrm{d}r} \left( 1+y^2\dot{r}^2-\frac{2GM(r)}{r} \right)\\
&-\frac{GM(r)}{r^2}-4\pi Gpr.
\end{split}
\end{equation}

In the presence of a cosmological constant $\Lambda$ with an equation of state like Eq. (\ref{eos}), this is slightly modified

\begin{equation}
\label{Eom2}
\begin{split}
 y\frac{d}{\mathrm{d}t}\left(y\frac{\mathrm{d}r}{\mathrm{d}t}\right)=&-\frac{1}{y}\frac{dp}{\mathrm{d}r}\left(1+y^2\dot{r}^2-\frac{8\pi G}{3}\left(\rho+\rho_{\Lambda}\right)r^2\right) \\ 
&-\frac{4\pi G}{3}\left(\rho-2\rho_{\Lambda}\right)r-{4\pi Gpr}.
\end{split}
\end{equation}

In case of equilibrium, Eq. (\ref{Eom}) and Eq. (\ref{Eom2}) reduce to the TOV equations with or without a cosmological constant:

\begin{eqnarray}
-p'&=&\frac{4 \pi G}{3} r \frac{\left(\rho+p\right)\left(\rho+3p\right)}{1-\frac{8 \pi G}{3}\rho r^2} \\
\label{TOV2}
-p'&=&\frac{4 \pi G}{3} r \frac{\left(\rho+p\right)\left(\rho+3p-2\rho_{\Lambda}\right)}{1-\frac{8 \pi G}{3}\left(\rho+\rho_{\Lambda}\right)r^2}.
\end{eqnarray}

If only small oscillations of the system around its equilibrium are considered, we can assume that $\dot{r}^2/c^2\ll 1$. Terms of this kind will be neglected in the following. After performing a Taylor-expansion up to the first post-Newtonian order $\mathcal{O}(\frac{1}{c^2})$ and inserting the zeroth-order expansion of the TOV-equation (Eq. (\ref{TOV2})), 

\begin{equation}
-p'=\frac{4\pi Gr}{3} \rho \left(\rho-2\rho_{\Lambda}\right)+ O\left(\frac{1}{c^2}\right),\\
\end{equation} 

we arrive at

\begin{equation}
\begin{split}
   \rho y^2 \ddot{r}=&-p'-\frac{4\pi Gr}{3}\rho\left(\rho-2\rho_{\Lambda}\right) \\
&-\left(\frac{4\pi Gr}{3}\left(\rho-2\rho_{\Lambda}\right)+4 \pi G \rho r\right)p \\ 
&- \left(\frac{32 \pi^2 G^2 r^3}{9}\rho \left(\rho^2-\rho \rho_{\Lambda}-2\rho_{\Lambda}^2\right) \right).
\end{split}
\end{equation}

Since the derivation of the virial theorem requires an integration over the spacetime volume element, the metric has to be expanded as well. In our case, spacetime is described by the TOV metric given by

\begin{equation}
\begin{split}
   ds^2=&-(1-AR^2)\left(\frac{3-\left(\frac{1-Ar^2}{1-AR^2}\right)^{1/2}f}{3-f}\right)^2 \mathrm{d}t^2+\frac{\mathrm{d}r^2}{1-Ar^2}\\
        &+r^2d\Omega^2
\end{split}
\end{equation}

where

\begin{equation*} 
 A=\frac{8 \pi G}{3} \left(\rho+\rho_{\Lambda}\right), \quad  f=1 - \frac{2\rho_{\Lambda}}{\rho}.
\end{equation*}

An expansion up to  $\mathcal{O}(1/c^2)$ leads to

\begin{equation}
\begin{split}
 ds^2\approx&-\left(1-2AR^2-\frac{f}{(3-f)}A \left(R^2-r^2\right)\right)\mathrm{d}t^2\\
&+\left(1+AR^2\right)\mathrm{d}r^2 + r^2d\Omega^2
\end{split}.
\end{equation}

The canonical volume form becomes

\begin{equation}
\begin{split}
 &\eta=\sqrt{-g} \ \mathrm{d}V_T\\ &=\sqrt{\left(\frac{1-AR^2}{1-Ar^2}\right)\left(\frac{3-\left(\frac{1-Ar^2}{1-AR^2}\right)^{1/2}f}{3-f}\right)^2} \mathrm{d}V_T \\
 &\approx\left(1+\frac{A}{2}\left(r^2-R^2\right)+\frac{f}{2(3-f)}A\left(r^2-R^2\right) \right)\mathrm{d}V_T 
\end{split}
\end{equation}

with $\mathrm{d}V_T=\mathrm{d}t \wedge \mathrm{d}V$ being the total volume element for a flat spacetime in spherical polar coordinates

\begin{equation}
 \mathrm{d}V_T=r^2\sin\theta \cdot \mathrm{d}t \wedge \mathrm{d}r \wedge d\theta \wedge d\phi.
\end{equation}
 
In the following, we will also apply the definition of the canonical volume form of the spacelike 3-hypersurface $\Sigma$ described by the spatial coordinates

 \begin{equation}
 \mathrm{d}V_\Sigma=\sqrt{\left. -g \right|_{\Sigma}} \ r^2\sin\theta \cdot  \mathrm{d}r \wedge d\theta \wedge d\phi.
\end{equation}

Taking the first spatial moment (multiplying with $r$ and integrating over the spatial volume) leads to the post-Newtonian version of Lagrange's identity (see \cite{collins_virial_1978}):

\begin{equation} 
\begin{split}
\int{\rho y^2 \ddot{r} r \mathrm{d}V_\Sigma}=-&\int{r p' \mathrm{d}V_\Sigma}-\int{\frac{4 \pi G r^2}{3}\left(\rho-2\rho_{\Lambda}\right)\rho \mathrm{d}V_\Sigma}\\
-&\int{\left(\frac{4\pi G r^2}{3}\left(\rho-2\rho_{\Lambda}\right)+4\pi G \rho r^2\right)p\mathrm{d}V_\Sigma} \\
-&\int{\frac{32\pi^2 G^2 r^4}{9}\left(\rho^2-\rho \rho_{\Lambda}-2\rho_{\Lambda}^2\right)\rho \mathrm{d}V_\Sigma}. 
\end{split}
\end{equation}

In analogy to the classical case, we interpret\footnote{$I_r$ is defined to be the relativistic generalization of the classical moment of inertia. For a homogeneous sphere it is classically defined by $I = \frac{1}{2} \int_V{\rho r^2 \mathrm{d}V}$. (see \cite{collins_virial_1978, maor_virialization_2005}}

\begin{eqnarray*} 
\frac{1}{2}\frac{d^2I_r}{\mathrm{d}t^2}&\equiv& \int{\rho y^2 \ddot{r} r \mathrm{d}V_\Sigma}\\
2T &\equiv& -\int{r p' \mathrm{d}V_\Sigma}.
\end{eqnarray*}

Using these definitions, Lagrange's identity becomes the familiar expression

\begin{equation}
\begin{split}
\label{virp}
\frac{1}{2} \frac{d^2I_r}{\mathrm{d}t^2}= &\ 2T-\int{\frac{4 \pi G r^2}{3}\left(\rho-2\rho_{\Lambda}\right)\rho \mathrm{d}V_\Sigma}\\
&-\int{\left(\frac{4\pi G r^2}{3}\left(\rho-\rho_{\Lambda}\right)+4\pi G \rho r^2\right)p\mathrm{d}V_\Sigma} \\ 
&-\int{\frac{32\pi^2 G^2 r^4}{9}\left(\rho^2-\rho \rho_{\Lambda}-2\rho_{\Lambda}^2\right)\rho \mathrm{d}V_\Sigma}. 
\end{split}
\end{equation}

Dropping the corrections in $1/c^2$, the classical version of Lagrange's identity is 

\begin{equation}
 \frac{1}{2}\frac{d^2I_{r}}{\mathrm{d}t^2}=2T+U_m-2U_\Lambda.
\end{equation}

Performing the time average will lead to the post-Newtonian virial theorem, because motions like oscillations around the equilibrium configuration are averaged out. Since we have to apply

\begin{equation}
 \lim_{T\rightarrow \infty}{\int_\mathrm{0}^T{\left(\ldots \right) e^a \mathrm{d}t}},
\end{equation}

the volume element of the averaged form changes $\mathrm{d}V_\Sigma=e^b\mathrm{d}V \rightarrow e^{a+b}\mathrm{d}V=\sqrt{-g}\mathrm{d}V$ while the time integration is performed.

Dropping all terms of $\mathcal{O}(1/c^4)$, the post-Newtonian virial theorem is:

\begin{equation}
\begin{split}
\label{vir_PN}
2T=&\int{\frac{4 \pi G r^2}{3}\left(\rho-2\rho_{\Lambda}\right)\rho \mathrm{d}V} \\
&+\int{\left(\frac{4\pi G r^2}{3}\left(\rho-2\rho_{\Lambda}\right)+4\pi G \rho r^2\right)p\mathrm{d}V} \ \text{(I)} \\ 
&+\int{\frac{32\pi^2 G^2 r^4}{9}\left(\rho^2-\rho \rho_{\Lambda}-2\rho_{\Lambda}^2\right)\rho \mathrm{d}V} \ \text{(II)}\\
&+\int{\frac{2\pi G r^2}{3}\left(\rho-2\rho_{\Lambda}\right)\left(1+\frac{f}{3-f}\right)A \left(r^2-R^2\right)}\rho \mathrm{d}V \ \text{(III)}.
\end{split}
\end{equation}

It can be seen that the correction terms contain 

\begin{setlength}{\leftmargini}{0.3cm}
\begin{itemize}
 \item
\textbf{pressure contributions} (I), since pressure acts as a source of gravity 
 \item
\textbf{backreaction terms} (II) between the fluid components and the geometry of space (due to the non-linearity of GR)
\item
\textbf{metric expansion terms} (III), since a non-vanishing energy-momentum tensor changes the metric (due to the field equations)

\end{itemize}
\end{setlength}

The potential energy given by Eq. (\ref{U}) can be expanded in the same way:

\begin{equation}
U=-4 \pi \rho  \int_\mathrm{0}^R{\left(\frac{A}{2}r^4+\frac{3}{8}A^2r^6\right)\mathrm{d}r}.
\end{equation}

Performing the angular integration for the kinetic energy expression leads to

\begin{equation}
\begin{split}
T=&\int_\mathrm{0}^R{\frac{8 \pi^2 G r^4}{3}\left(\rho-2\rho_{\Lambda}\right)\rho \mathrm{d}r}\\
&+\int_\mathrm{0}^R{\left(\frac{8\pi^2 G r^4}{3}\left(\rho-2\rho_{\Lambda}\right)+8 \pi^2 G \rho r^4\right)p\mathrm{d}r} \\ 
&+\int_\mathrm{0}^R{\frac{64\pi^3 G^2 r^6}{9}\left(\rho^2-\rho \rho_{\Lambda}-2\rho_{\Lambda}^2\right)\rho \mathrm{d}r}\\
&+\int{\frac{32\pi^3 G^2 r^4}{9}\left(\rho-2\rho_{\Lambda}\right)\left(\rho+\rho_{\Lambda} \right) \rho \left(1+\frac{f}{3-f}\right)}\\ 
& \quad \left(r^2-R^2\right) \mathrm{d}r.
\end{split}
\end{equation}

Now we rewrite some variables\footnote{$R_\mathrm{ta}$ is again calculated using Eq. (\ref{R_ta}).} 
 
\begin{equation*} 
r = y\cdot R_\mathrm{ta}, \quad R_\mathrm{vir} = y_\mathrm{vir} \cdot R_\mathrm{ta}
\end{equation*}

and the virialization equation becomes

\begin{equation}
\label{virialization_PN}
\left[T+U\right]_\mathrm{vir}=U_\mathrm{ta}
\end{equation}

with the terms

\begin{equation}
U_\mathrm{vir}=-4 \pi \rho_\mathrm{vir}  \int_\mathrm{0}^{y_\mathrm{vir}}{\left(\frac{A_\mathrm{vir}}{2}y^4+\frac{3}{8}A_\mathrm{vir}^2y^6 R_\mathrm{ta}^2\right)R_\mathrm{ta}^5dy}
\end{equation}

\begin{equation}
U_\mathrm{ta}=-4 \pi \rho_\mathrm{ta}  \int_\mathrm{0}^1{\left(\frac{A_\mathrm{ta}}{2}y^4+\frac{3}{8}A_\mathrm{ta}^2y^6 R_\mathrm{ta}^2\right)R_\mathrm{ta}^5dy}
\end{equation}

and 

\begin{equation}
\label{T_PN}
\begin{split}
T_\mathrm{vir}&= \int_\mathrm{0}^{y_\mathrm{vir}}{\frac{8 \pi^2 G y^4 R_\mathrm{ta}^5}{3}\left(\rho_\mathrm{vir}-2\rho_{\Lambda}\right)\rho_\mathrm{vir} dy}\\ 
&+\int_\mathrm{0}^{y_\mathrm{vir}}{\left(\frac{8\pi^2 G y^4 R_\mathrm{ta}^5}{3}\left(\rho_\mathrm{vir}-2\rho_{\Lambda}\right)+8 \pi^2 G \rho_\mathrm{vir} y^4 R_\mathrm{ta}^5\right)p_\mathrm{vir}dy} \\ 
&+\int_\mathrm{0}^{y_\mathrm{vir}}{\frac{64\pi^3 G^2 y^6 R_\mathrm{ta}^7}{9}\left(\rho_\mathrm{vir}^2-\rho_\mathrm{vir} \rho_{\Lambda}-2\rho_{\Lambda}^2\right)\rho_\mathrm{vir} dy}\\ 
&+\int_\mathrm{0}^{y_\mathrm{vir}}{\frac{32\pi^3 G^2 y^4R_\mathrm{ta}^7}{9}\left(\rho_\mathrm{vir}^3-\rho_\mathrm{vir}^2\rho_{\Lambda}-2\rho_\mathrm{vir}\rho_{\Lambda}^2\right)}\\
& \quad \left(1+\frac{f}{3-f}\right)\left(y^2-y_\mathrm{vir}^2\right) dy.
\end{split}
\end{equation} \\

In addition, we have applied the following definitions: 

\begin{equation*} 
\begin{split}
&\rho_\mathrm{vir} = \frac{3M}{4\pi y_\mathrm{vir}^3 R_\mathrm{ta}^3}, \quad \rho_\mathrm{ta} = \frac{3M}{4 \pi R_\mathrm{ta}^3}, \quad A_\mathrm{vir}=\frac{8\pi G}{3 } \left(\rho_\mathrm{vir}+\rho_{\Lambda}\right),\\
& A_\mathrm{ta}=\frac{8\pi G}{3 } \left(\rho_\mathrm{ta}+\rho_{\Lambda}\right).
\end{split}
\end{equation*}

As for the fully relativistic version, Eq. (\ref{virialization_PN}) has to be solved for $y_\mathrm{vir}$ numerically.

\section{The relativistic formalism and dynamical DE models} \label{sec:dyn_DE}

Even though we have spent some effort to generalize our method to dynamical DE models, certain problems occur which will be described in the following:\\

Consider a two component fluid described by

\begin{eqnarray}
 T_{\mu\nu}^{(m)}&=& \left(\rho+p\right) u_{\mu} u_{\nu} + p g_{\mu\nu} \\
 T_{\mu\nu}^{(Q)}&=& \left(\rho_Q+p_Q\right) u_{\mu} u_{\nu} + p_Q g_{\mu\nu} \\ \nonumber
\\
T_{\mu\nu} &=& T_{\mu\nu}^{(m)} + T_{\mu\nu}^{(Q)}
\end{eqnarray}

where the densities $\rho$ and $\rho_Q$ are assumed to be constant and the quintessence component has an equation of state $p_Q=w\rho_Q$ with constant $w$. Energy-momentum conservation is separately fulfilled for each fluid component

\begin{eqnarray}
 \nabla_{\mu}T^{\mu\nu, m}&=&0 \\
 \nabla_{\mu}T^{\mu\nu, Q}&=&0.
\end{eqnarray}

The static, spherically symmetric field equations for this set-up are

\begin{eqnarray} \nonumber
 G_{\mu\nu}=8\pi G T_{\mu\nu}\\ \nonumber \\
\label{w_feqs 1}
\frac{1}{r^2}-e^{-2b}\left(\frac{1}{r^2}-\frac{2b'}{r}\right) &=& 8\pi G \left(\rho + \rho_Q \right) \\
\label{w_feqs 2}
-\frac{1}{r^2}+e^{-2b}\left(\frac{1}{r^2}+\frac{2a'}{r}\right) &=& 8\pi G \left( p + w\rho_Q \right)\\
\label{w_feqs 3}
e^{-2b}\left(a''-a'b'+a'^2+\frac{a'-b'}{r}\right) &=& 8\pi G \left( p+ w \rho_Q \right).
\end{eqnarray}

It turns out that, in case of $w\neq -1$, Eqs. (\ref{w_feqs 1}) - (\ref{w_feqs 3}) are no longer consistently solvable and lead to contradictory solutions (see Appendix \ref{app_C} for a detailed proof). This means that a model based on

\begin{setlength}{\leftmargini}{0.3cm}
 \begin{itemize}
  \item \textbf{staticity}
  \item \textbf{spherical symmetry}
  \item \textbf{matter model} (two component fluid consisting of clustering dark matter and homogeneous DE with equation of state $p_Q=w \rho_Q$)
 \end{itemize}
\end{setlength}

does not lead to a consistent description. Static, spherically symmetric solutions with $w \neq -1$ can therefore only be given approximately. In order to achieve exact solutions in the general case, we need to drop at least one of the model assumptions. Since  spherical symmetry is dictated by the spherical collapse model and the matter model and we want to fix the matter model, we can only stick to time-dependent problems and drop staticity.\footnote{It has to be mentioned for completeness that there exist static exterior solutions describing a Schwarzschild black hole surrounded by a Quintessence fluid (see \cite{fernando_schwarzschild_2012, kiselev_quintessence_2003}.). However, in these cases  $\rho_Q$ is constrained to be radial-dependent: $\rho_Q(r)\sim r^{-3(1+w)}$.} \\

A physical explanation for this constraint on $w$ is local energy-momentum conservation. Since we assume that it has to be valid for each component separately and the field equations are constructed in a way that energy and momentum are locally conserved, a static quintessence component with $w \neq -1$ violates this requirement.\\

Consider a perfect fluid with  $T_{\mu\nu}=(\rho_Q+p_Q)u_{\mu}u_{\nu}+p_Q g_{\mu\nu}$ which has to obey local energy-momentum conservation

\begin{equation}
 \nabla_\mu T^{\mu\nu, Q} =0. 
\end{equation}

Projecting this onto the 3-space perpendicular to the mean-fluid flow, we obtain the relativistic Euler equation

\begin{equation}
 \left(g_{ \alpha\nu} + u_{\alpha}u_{\nu}\right)\nabla_\mu T^{\mu\nu,Q} =0 
\end{equation}

which leads to

\begin{equation}
\label{rel_Euler}
 \left(\rho_Q+p_Q\right) \nabla_u u = -\textnormal{grad} \ p_Q -u \nabla_u p_Q.
\end{equation}\\

In case of a static configuration ($u=(1, 0, 0, 0)$, $p_Q(r,t)= p_Q(r)$) and the metric ansatz from Eq. \ref{metric2}, we are left with
 
\begin{equation}
 - p_Q' = a'\left(\rho_Q+p_Q\right)
\end{equation}

which becomes for the given equation of state $p_Q = w \rho_Q$:

\begin{equation}
 0 = a' \rho_Q \left(1+w\right).
\end{equation}

Since $a'$ is non-zero in general, we have to require $w=-1$ in order to satisfy local energy-momentum conservation. Thus, a quintessence fluid with constant density and $w\neq -1$ violates local energy-momentum conservation in case of a static configuration.\\ 

Energy and momentum are certainly conserved for a time-dependent DE density which scales like $\rho_Q = \rho_Q^{(0)} a^{-3(1+w)}$\footnote{Of course, this is only true for constant equation-of-state parameter $w$. In the general case, $\rho_Q$ evolves like $\rho_Q = \rho_Q^{(0)} g(a)$ with $g(a)=\exp\left(-3 \int_1^a{(1+w(a'))d\ln{a'}}\right)$ (see for example \cite{bartelmann_non-linear_2006})}. In this case, the only time-independent DE density is the cosmological constant representing the only possible static, quintessence fluid configuration with constant equation of state that satisfies local energy-momentum conservation. In our approach, we restricted ourselves to homogeneous DE that evolves independently from the matter component. In case of clustering DE (see \cite{basilakos_virialization_2007, maor_spherical_2007, maor_virialization_2005, nunes_structure_2006}) or even models considering interactions between the two fluids (see \cite{manera_cluster_2006, wintergerst_clarifying_2010}), our formalism will significantly change and might allow an application to these models.

\section{Results}\label{sec:res}

\begin{figure*}[ht]
\centering
\begin{tabular}{lr}
 \resizebox{90mm}{!}{\rotatebox{270}{\includegraphics{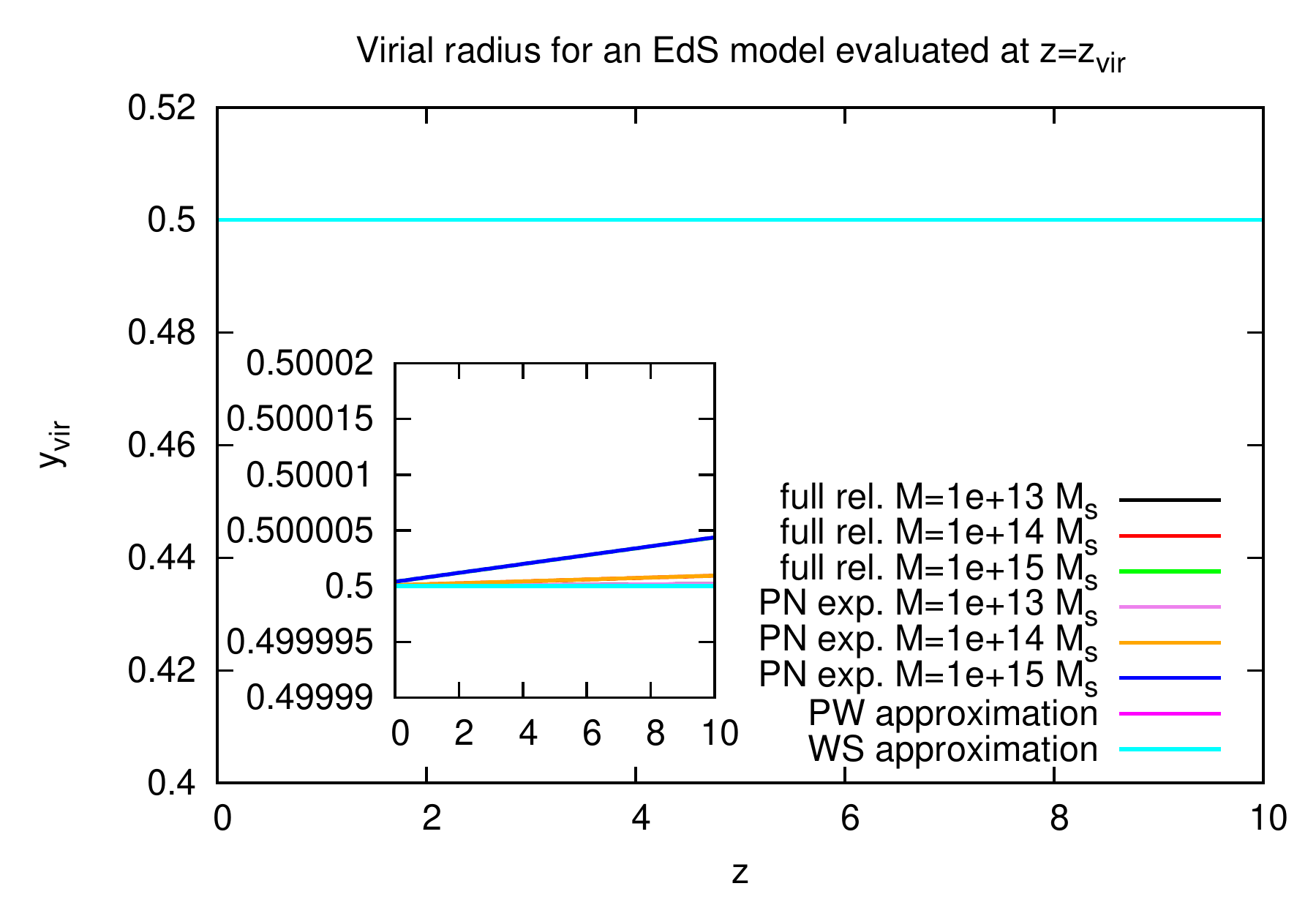}}} &
 \resizebox{90mm}{!}{\rotatebox{270}{\includegraphics{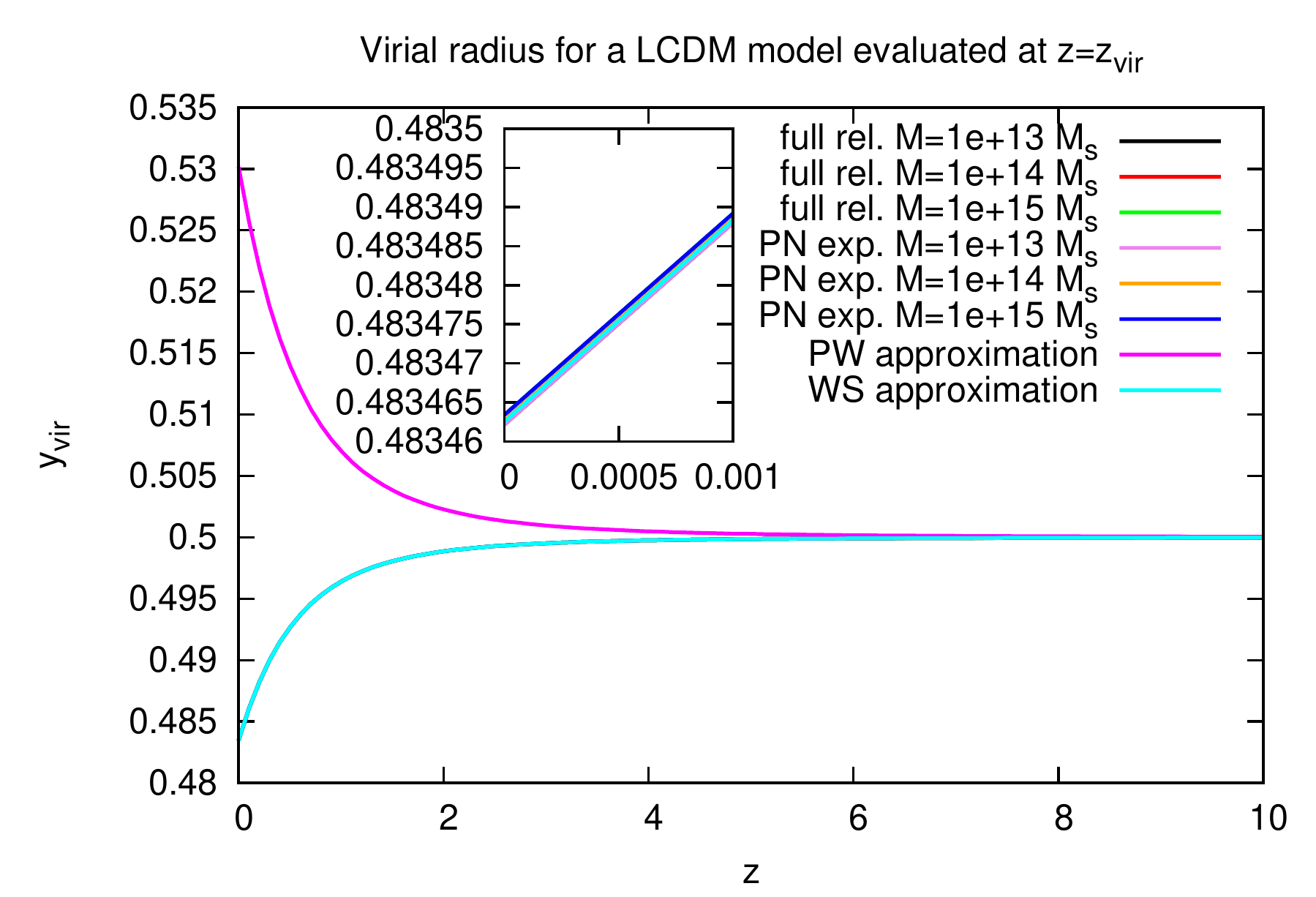}}} \\
 \resizebox{90mm}{!}{\rotatebox{270}{\includegraphics{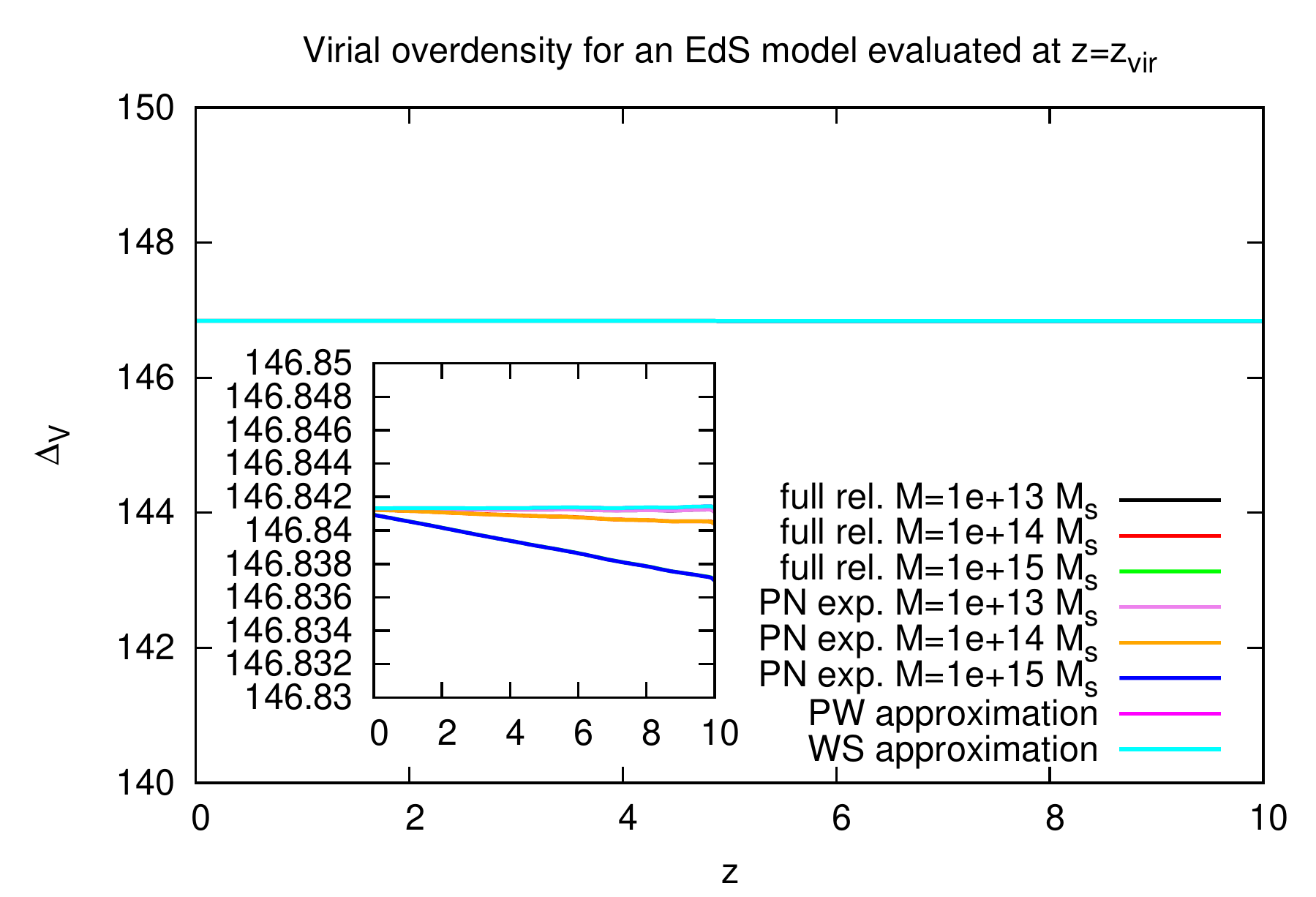}}} &
 \resizebox{90mm}{!}{\rotatebox{270}{\includegraphics{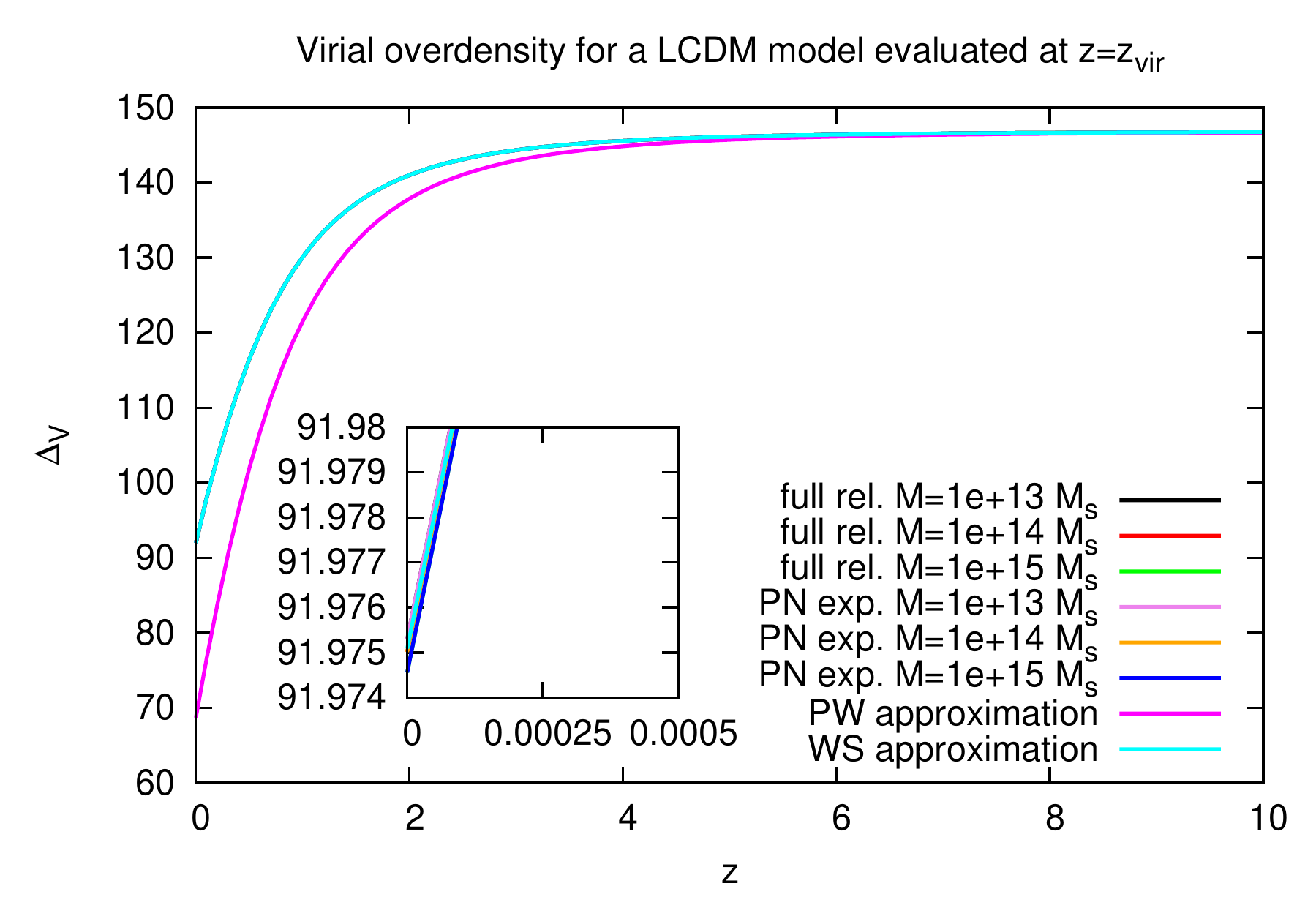}}} \\
\end{tabular}
\caption{Virial radius and virial overdensity obtained by the relativistic virialization equation and its post-Newtonian expansion as function of collapse redshift for EdS and $\Lambda$CDM cosmologies. The upper panels show the virial radius for three different halo masses obtained by solving the full relativistic virialization equation (Eqs. (\ref{virialization_eq})-(\ref{pot_ta})) and its post-Newtonian expansion (Eqs. (\ref{virialization_PN}) - (\ref{T_PN})) using the two cosmological models. The lower panels show the corresponding virial overdensity obtained by Eq. (\ref{Delta_V}). The classical solutions of Wang \& Steinhardt (1998) (cyan line) and Wang (2006) (purple line) are plotted for reference. The region close to $z_\mathrm{c}=0$ has been scaled up to sufficiently small redshifts in order to illustrate the dependence on the halo mass.} 
\label{relat_vir}
\end{figure*}

The relativistic virialization equation and its post Newtonian expansion are solved for the virial radius. Fig. (\ref{relat_vir}) shows the virial radius and overdensity as a function of the collapse redshift $z_{\mathrm{c}}$ for three different halo masses in an EdS and $\Lambda$CDM cosmology. All quantities at virialization are evaluated at the exact virial redshift $z_{\mathrm{vir}}$. It is a common approximation in the spherical collapse model to insert the potential energy evaluated at the collapse redshift. As already investigated analytically by Lee \& Ng (2010) (see \cite{lee_spherical_2010}) the result of the virial overdensity changes significantly\footnote{In the EdS case $\Delta_V(z_\mathrm{c})\sim 178$ changes into $\Delta_V(z_\mathrm{vir})\sim 146$.} by inserting the exact virial redshift instead. We have developed and applied an iterative method to obtain $z_\mathrm{vir}$ numerically and the results of Lee \& Ng are nicely reproduced. The derivation is postponed to Appendix \ref{app_D}.\\

It can be seen clearly in both figures that the Wang-Steinhardt-limit is precisely recovered. This is expected, because the expressions for the potential energy and the kinetic energy derived in Sect. \ref{sec:vir_rel} contain the WS solution as limit to zeroth order. Since the halo mass does no longer cancel out naturally on both sides of the virialization equation, the spherical collapse becomes mass-dependent. Therefore each result is plotted for three different masses, namely  $10^{13} M_{\odot}, \ 10^{14} M_\odot \ \text{and } \ 10^{15} M_\odot$. Nevertheless, it can be seen that the results for $M \ = \ 10^{13} M_\odot$ and $M \ = \ 10^{15} M_\odot$ differ from each other by only $10^{-3}\%$.

Since, averaged over sufficiently large timescales, the final state of a virialized cluster is static\footnote{Oscillations around the virial radius can be expected to average out.}, it can be considered as a homogeneous, static perfect-fluid sphere such that the Tolman-Oppenheimer-Volkoff solution can be applied. We have constructed the virialization equation based on the TOV solution instead of using the classical approach from Friedmann's equations. A few points have to be discussed concerning this method:

\begin{setlength}{\leftmargini}{0.3cm}
\begin{itemize}
 \item As an important approximation, we have assumed that the Killing vector field $K$ of the FLRW universe is time-like on halo scales. Since the final state is static anyway, this is most relevant for the turn-around which is described by a static solution as well. As shown in Appendix \ref{app_A}, the space-like component of $K$ is by at least two orders of magnitude smaller than the time-like one. Looking at the small corrections in first post-Newtonian order in Tab. \ref{vir_PN_contr} being five orders of magnitude smaller than the classical term, it remains an open question, whether a time-dependent approach that does not obey that approximation, would have a non-negligible effect on the results.
\item As can be seen in Tab. \ref{vir_PN_contr}, the normalized contributions to the first post-Newtonian order are of about  $10^{-3}\%$ being almost independent of the type of contribution. This can be expected due to a simple estimate. Let us assume a typical massive  galaxy cluster with a mass of $10^{15} M_\odot$ and a virial radius of $1$ Mpc. The gravitational potential $\phi/c^2$ being the ratio of its Schwarzschild radius and virial radius has the value

\begin{equation}
 \frac{\phi}{c^2} = \frac{GM}{c^2 R_\mathrm{vir}}  \sim 5\cdot  10^{-5}.
\end{equation}
 
Thus, the post Newtonian terms which are of the order $(\phi/c^2)^2$ must have absolute values of about $10^{-10}$ which corresponds to relative contribution of $10^{-5}$ ($10^{-3}\%$) with respect to the classical term. 
\end{itemize}
\end{setlength}

The key question remains why the relativistic calculation reduces to the WS limit instead of the result of PW. The ansatz of a static, spherically symmetric metric reduces Einstein's field equations to a coupled system of three ordinary differential equations with respect to the radius $r$ (see Eqs. (\ref{feqs 1}) - (\ref{feqs 3})). Eq. (\ref{feqs 1}) is already decoupled from Eqs. (\ref{feqs 2}) and (\ref{feqs 3}) such that the $rr$-component of the metric is constrained to be

\begin{equation}
\label{g_rr}
 g_{rr}= e^{2b} = \frac{1}{1-Ar^2}, \quad A=\frac{8\pi G}{3} \left(\rho+\rho_{\Lambda}\right).
\end{equation}

It has to be mentioned that Eq. (\ref{g_rr}) does not contain any pressure term. Straumann's self energy expression derived in Appendix \ref{app_B} is based on the number density $n(r)$ of dark matter particles that is integrated over the covariant volume element restricted to the hypersurface $\Sigma$ spanned by the spatial coordinates: 

\begin{equation}
 \left. \eta \right|_{\Sigma}= \sqrt{\left. -g\right|_{\Sigma}} \mathrm{d}r d\theta d\phi = \frac{r^2 \sin{\theta}}{\sqrt{1-Ar^2}} \mathrm{d}r d\theta d\phi.
\end{equation}

Finally, the result becomes

\begin{equation}
\begin{split}
    U &= \int_\mathrm{0}^R {4 \pi r^2 \rho \left(1- \frac{1}{\sqrt{1-Ar^2}} \right) \mathrm{d}r} \\
      &\approx -\frac{3}{5} \frac{G M^2}{R} -\frac{4 \pi G}{5} M \rho_{\Lambda} R^2 + \mathcal{O}(A^2)  
\end{split}
\end{equation}

which perfectly reproduces the potential energy used by Wang \& Steinhardt in their model. \\

The TOV equation for the pressure profile $p(r)$ is a generalization of the Newtonian hydrostatic equation. Pressure is contained as an additional source of gravity, but the post-Newtonian expansion of the resulting virial theorem in Eq. (\ref{vir_PN}) shows clearly that  the limit of WS is reproduced to zeroth order. Horellou \& Berge have shown in 2005 (see \cite{horellou_dark_2005}) that, from a classical point of view, the Wang-Steinhardt-solution is exactly valid in the $\Lambda$CDM case. This means that for the classical self-energy expression of a matter sphere in a $\Lambda$CDM background inserted into the classical virialization equation (see Sect. \ref{sec:class}) the solution of Wang \& Steinhardt is recovered.\\

To conclude this section we can state the following: Based on the assumptions of a static, spherically-symmetric spacetime and separately fulfilled energy-momentum conservation equations for matter and cosmological constant\footnote{The cosmological constant fluid trivially fulfills energy momentum conservation, since the continuity equation reduces to $\dot{\rho}_{\Lambda}=0$.}, the field equations and the resulting TOV equation provide a relativistic virial theorem that contains the Wang-Steinhardt-result as limit to zeroth order. It remains an open question whether a non-static approach to relativistic gravitational collapse that might also be adapted to dynamical DE models would still recover this result.

\begin{table}
\centering 
\begin{tabular}{rl}
\hline
\hline
 \multicolumn{2}{c}{\textbf{Einstein de-Sitter:}} \\
\hline
 pressure term & $1.274132394 \cdot 10^{-5}$ \\
 backreaction term & $3.18533352 \cdot 10^{-5}$ \\
 metric expansion & $9.5560225 \cdot 10^{-6}$\\
 \hline
 \hline
 \multicolumn{2}{c}{\textbf{$\Lambda$CDM:}}\\
 \hline
 pressure term & $1.2893437 \cdot 10^{-5}$ \\
 backreaction term & $2.7275899168 \cdot 10^{-5}$ \\
 metric expansion & $7.073241606 \cdot 10^{-6}$\\
 \hline
 \hline
\end{tabular}
\caption[Post-Newtonian contributions]{The three relative post-Newtonian contribution terms with respect to the classical Newtonian term are considered in EdS and $\Lambda$CDM cosmology for $z_{\mathrm{c}}=0$ and $M=10^{15}M_{\odot}$.}
\label{vir_PN_contr}
\end{table}

\section{Remarks on the required pressure profile}

As already mentioned by Oppenheimer and Snyder in 1939 (see \cite{oppenheimer_continued_1939}), there does not exist any static, spherically-symmetric solution of the field equations with vanishing pressure. If no positive pressure profile is present, nothing will prevent the spherical object from collapsing into a singularity. Therefore, in the classical spherical collapse, the virialization condition is introduced to define an equilibrium. In our case, gravity forces the system to build up a pressure profile to prevent the sphere from collapsing into a singularity. In fact, the process of virialization must convert ordered motion from the collapse into unordered motion and the kinetic energy associated with the unordered motion corresponds to pressure. Thus, an equilibrium is reached by a positive pressure profile that can be related directly to the mean kinetic energy of the system (see Eq. (\ref{pres})). This might be a first step towards a more fundamental theory of virialization avoiding of the interpretation of an "enforced" equilibrium.\\
The non-linear density evolution equation (see Eq. (\ref{nlm})) is still based on extended Newtonian theory.\footnote{Velocities and gravitational potentials are assumed to be small compared to $c^2$ and pressure is included as a source of gravity.} In order to achieve full consistency, a relativistic evolution equation for either the density or the radius is recommended. Nevertheless, these equations will retain their full validity if dark matter is assumed to be pressureless during the evolution and that the pressure profile is built up instantaneously at virialization\footnote{Nevertheless, we have to admit that the assumption of an instantaneously appearing pressure profile at virial redshift is a highly idealized concept.}. The appearance of a pressure profile in this context is a direct consequence of the spherical collapse model itself. Strictly speaking, virialization is a tool in the spherical collapse model in order to achieve an equilibrium state. The pressure is directly related to the mean kinetic energy of the system which is again related to the potential energy by the virial theorem. Since the last two quantities cannot vanish in general, neither can the pressure. Thus, the virial theorem which is combined with, but in no way related to the non-linear density evolution requires a pressure profile, even if the latter is based on cold dark matter.

\section{Conclusion and outlook}

We propose a way to set up a fully relativistic method to obtain the virial radius and the virial overdensity for the EdS and $\Lambda$CDM cosmology. Within the assumption of an approximately time-like Killing vector field of the FLRW metric, static solutions for perfect fluid spheres in general relativity (namely the Tolman-Oppenheimer -Volkoff equation) have been successfully applied to extend the virial theorem in a consistent manner. The result has been inserted into the virialization equation which can be solved for the virial radius to find the corresponding virial overdensity. It turns out that the solution of Wang \& Steinhardt (1998) (\cite{wang_cluster_1998}) for the virial radius in a $\Lambda$CDM cosmology is almost perfectly reproduced by our formalism which can also be shown analytically by performing the weak field limit. The first order post-Newtonian expansion has been investigated and the leading order corrections have been worked out and calculated numerically. We found out that they have a relative contribution of $10^{-3}\%$ with respect to the classical term which is of very small, but expected size. Although these corrections are of limited astrophysical interest, the concept itself is a small step towards a more fundamental understanding of virialization of spherical halos in the presence of DE. In addition, an iterative method has been set up to calculate the exact virial redshift numerically. The results of Lee \& Ng (2010) (\cite{lee_spherical_2010}) are reproduced extremely well.\\

 Naturally, galaxies and galaxy clusters are far more complex than homogeneous and isotropic spheres, but the spherical collapse model provides a very simple semi-analytic method that already suffices to estimate important parameters like the virial radius and virial overdensity. The process of virialization itself is an additional condition that has been introduced to prevent a spherical overdensity of pressureless dark matter from collapsing into a singularity. A pressure profile, which a relaxed continuous spherical object must obey due to general relativity, can provide new insight into the process of virialization itself.\\

There are certainly some topics this paper cannot address, because they are far beyond its scope. Although the spherical collapse models are powerful tools to obtain estimates of the evolution of structures in the universe, they are limited by their simplicity. In particular, the fact that the spherical overdensity is in no way embedded continuously into the background Friedmann universe is still very idealized and dissatisfying. Secondly, DE cannot be described yet in a self-consistent way with general relativity, since local energy-momentum-conservation, which is required by the theory, is not fulfilled and a coordinate representation of the two fluids is missing. Approaches based on the Lema\^{i}tre-Tolman-Bondi models (see \cite{bondi_spherically_1999, marra_exact_2012, pereira_evolution_2010, romano_constructing_2011, tolman_effect_1997, valkenburg_exact_2011}) as well as the presented work of Misner and Sharp (1964) (see \cite{misner_relativistic_1964}) are promising candidates for following investigations.

\bibliography{references/references_abbrev}

\appendix

\section{Approximately time-like Killing vector field of FLRW spacetime on halo scales}\label{app_A}

Let us start with the FLRW-spacetime given by the metric

\begin{equation}
\label{metric}
 ds^2=-\mathrm{d}t^2+\frac{a^2(t)}{\left(1+\frac{kr^2}{4}\right)^2} \left(\mathrm{d}r^2+r^2d\Omega^2\right).
\end{equation}

This model is based on isotropy and homogeneity of the 3-dimensional spacelike hypersurfaces describing a space of constant curvature $k$.  \\

Isotropy requires the existence of a coordinate frame in which spatial rotations are isometries. Given that frame, the most general ansatz for a Killing vector field of Eq. (\ref{metric}) is:

\begin{equation}
\label{Killing}
 K=A(r,t)\partial_t+B(r,t)\partial_r
\end{equation}

with $A$ and $B$ being arbitrary scalar functions of radius and time.\\

Eq. (\ref{Killing}) has to be inserted into the Killing equation to obtain any relation between $A$ and $B$.

\begin{equation} 
\left(\mathcal{L}_K g\right)_{\mu\nu}=0. 
\end{equation}

The Lie derivative of a rank $(0,2)$ tensor can be written as 

\begin{equation}
\label{Killing_eq}
 K^{\lambda}\partial_\lambda g_{\mu\nu}+g_{\lambda \nu}\partial_\mu K^{\lambda}+g_{\mu\lambda}\partial_\nu K^{\lambda}=0.
\end{equation}

Inserting Eq. (\ref{Killing}) into Eq. (\ref{Killing_eq}) the following three constraints can be obtained:

\begin{eqnarray}
\label{Killing1}
 \left(\mathcal{L}_K g\right)_{00}&:& 2 \dot{A}=0\\ 
\label{Killing2}
 \left(\mathcal{L}_K g\right)_{11}&:& \left(2 \dot{a}A+2aB'\right) \left(1+\frac{k}{4}r^2\right)-kra B =0\\
\label{Killing3}
 \left(\mathcal{L}_K g\right)_{22}&:& \left(2 \dot{a}Ar^2+2a rB\right) \left(1+\frac{k}{4}r^2\right)\\
                                  &-&kr^3aB =0 \\ 
\label{Killing4}
 \left(\mathcal{L}_K g\right)_{01}&:& A' = \frac{a^2}{(1+\frac{kr^2}{4})^2} \dot{B} \\ \nonumber \\
\label{Killing5}
 \left(\mathcal{L}_K g\right)_{33} &=& \sin^2\theta \left(\mathcal{L}_K g\right)_{22}=0
\end{eqnarray}
 
Eq. (\ref{Killing1}) implies that $A$ has no time-dependence, thus 

\begin{equation}
 A(r,t)=A(r).
\end{equation}

If Eq. (\ref{Killing2}) and Eq. (\ref{Killing3}) are combined, we can obtain the differential equation in $B$

\begin{equation}
\label{Killing6}
 B'=\frac{B}{r}.
\end{equation}

This can be inserted into Eq. (\ref{Killing2}) to obtain a relation between $A$ and $B$

\begin{equation}
\label{relation}
 A=-\frac{B}{rH} \left[1-\frac{\frac{kr^2}{2}}{1+\frac{kr^2}{4}}\right].
\end{equation}

It has to mentioned for completeness that Eq. (\ref{relation}) has to be consistent with Eqs. (\ref{Killing1}) and (\ref{Killing4}). Calculating the derivative of Eq. (\ref{relation}) with respect to $r$ and equating this with Eq. (\ref{Killing4}) leads to an additional constraint for $B$: 

\begin{equation}
\begin{split}
& A' = \frac{B}{rH} \frac{kr}{\left(1+\frac{kr^2}{4}\right)^2} = \frac{a^2 \dot{B}}{\left(1+\frac{kr^2}{4}\right)^2}\\
& \Rightarrow \dot{B} = \frac{kB}{a^2H}.
\end{split}
\end{equation}

Let us now apply the following approximation:\\

 Since the spherical overdense region is of the scale of $1-10$ Mpc which is small compared the Hubble radius $R_H \sim \frac{1}{\sqrt{k}}\sim 1-3$ Gpc, it can be assumed that   

\begin{equation}
 r^2 \ll \frac{1}{k}
\end{equation}

which implies

\begin{equation}
 \frac{\frac{kr^2}{2}}{1+\frac{kr^2}{4}} \ll 1.
\end{equation}

Thus, we can write approximately for Eq. (\ref{relation}) (adding the factor $c$ again):

\begin{equation}
\begin{split}
  A&\approx-B\frac{c}{Hr}\\
 &=-B \left(\frac{c}{H_\mathrm{0}r}\right)\cdot \frac{1}{E(a)}\\
 &\approx -B \frac{c}{H_\mathrm{0} r}=  -B \frac{R_H }{r}
\end{split}
\end{equation}

because $E(a)=\sqrt{\Omega_m^{(0)}a^{-3}+\Omega_\Lambda^{(0)}+\Omega_K^{(0)}a^{-2}} \sim \mathcal{O}(1)$ (within the range of parameters we consider) and $R_H=c/H_\mathrm{0}$.\\

This implies

\begin{equation}
 \frac{B}{A} = \frac{r}{ R_H}.
\end{equation}

In our considered spherical collapse model, parameters are given within the following range
 
\begin{eqnarray*} 
 a_c, a_\mathrm{ta}  &\sim& \mathcal{O}(1) - \mathcal{O}(10^{-1}) \\
 r &\sim& 1 - 10 \text{Mpc} \\ 
 R_H &\sim& 3 \text{Gpc}. \\ 
 \end{eqnarray*}

Inserting this, we end up with

\begin{equation}
\label{AB}
 \frac{|B|}{|A|} \lesssim 3.3 \cdot 10^{-3} \approx 0.3 \%
\end{equation}

which means that the Killing vector field is approximately time-like:

\begin{equation}
 K \approx A(r) \partial_t.
\end{equation}

\section{Derivation of the relativistic potential energy}\label{app_B}

The following derivation is essentially taken from N. Straumann (\cite{straumann_general_2004}) and will be only slightly modified for our purposes:\\

 Let us consider a particle representation of dark matter and compare the total mass M of the spherical halo with the rest mass of the gravitationally interacting dark matter particles. The total rest mass of all DM particles is given by  

\begin{equation}
 M_\mathrm{0}=N m_N
\end{equation}
   
with $N$ being the total number of particles and $m_N$ the rest mass of a single particle. Let us define a current density $J$ as a one-form such that we can express the number of particles via a surface integral: 

\begin{equation}
N=\int_{t=const} \ast J. 
\end{equation}

$J$ can be expressed as

\begin{equation}
\begin{split}
  &J = J_{\mu}\theta^{\mu} \\
\Rightarrow & \ast J = J_{\mu} \ \ast\theta^{\mu} = J_{\mu}\eta^{\mu} \quad \text{with} \quad \eta^{\mu} = \ast\theta^{\mu}
\end{split}
\end{equation}

with respect to an arbitrary dual basis $\left\{\theta^{\mu}\right\}$.

We transform the integral by evaluating the Hodge dual explicitly:\footnote{Since we consider a static configuration we can find a coordinate frame in which the $J_i$ components vanish. We will assume that in the following without loss of generality.}

\begin{equation}
 \begin{split}
  \int_{t=const} \ast J &= \int_{t=const}{J_{\mu} \eta^{\mu}} \\
                    &= \int_{t=const}{J_{0} \eta^{0}} \\
                    &= \int_{t=const}{J_{0} \ast\theta^{0}}. \\ 
 \end{split}
\end{equation}

Assuming the metric ansatz given in Eq. (\ref{metric2}) and defining the dual basis like
\begin{equation}
 \theta^0 = e^a \mathrm{d}t, \quad \theta^1 = e^b \mathrm{d}r, \quad  \theta^2 = r^2 d\theta, \quad \theta^3 = r^2 \sin^2\theta d\phi.
\end{equation}

$\ast\theta^{0}$ can be evaluated:

\begin{equation}
 \begin{split}
  \ast\theta^0 &= e^a \ast \mathrm{d}t \\
               &= e^{b} r^2 \sin{\theta} \mathrm{d}r \wedge d\theta \wedge d\phi \\
               &= \theta^1 \wedge \theta^2 \wedge \theta^3.
 \end{split}
\end{equation}

Since, in a static configuration, the $\theta^{\mu}$ are an orthonormal system, the above result can be expected.\\

Thus, we are left with

\begin{equation}
\label{part_number}
 \begin{split}
  N &= \int_{t=const} {J_\mathrm{0} \theta^1 \wedge \theta^2 \wedge \theta^3} \\
    &= \int_{t=const} {J_\mathrm{0} e^b r^2 \sin{\theta} \mathrm{d}r \wedge d\theta \wedge d\phi}\\
    &= \int_\mathrm{0}^R {4 \pi r^2 J_\mathrm{0} e^b \mathrm{d}r}.
 \end{split}
\end{equation}

The number density $n(r)$ can be obtained by projection of $J$ onto the four velocity $u^\mu$ being $(1, 0, 0, 0)^T$ in the chosen coordinate frame: 

\begin{equation}
 n(r)= -u^\mu J_\mu = J_\mathrm{0}
\end{equation}

such that we get 

\begin{equation}
 N= \int_\mathrm{0}^R{4 \pi r^2 n(r) e^{b(r)} \mathrm{d}r}.
\end{equation}

We can define the proper DM energy density (total energy density with subtracted particle rest energy density)

\begin{equation}
\label{epsilon}
 \epsilon(r)= \rho(r)-m_N n(r)
\end{equation}

which corresponds to an intuitive proper internal energy of

\begin{equation}
 E = M-M_\mathrm{0} = M - N m_N.
\end{equation}

The proper internal energy can be decomposed into a total kinetic and a total potential energy of the system such that $T+V=M-M_\mathrm{0}$. Let us insert the integral for the particle number given by Eq. (\ref{part_number})

\begin{equation}
\label{rest}
\begin{split}
  m_N N &= \int_\mathrm{0}^R {4 \pi r^2 e^b m_N n(r) \mathrm{d}r}\\
        &= \int_\mathrm{0}^R {\left(\rho(r) - \epsilon(r)\right)4 \pi r^2 e^b \mathrm{d}r}\\
        &= M-T-V \\
        &= \int_\mathrm{0}^R {4\pi r^2 \rho(r) \mathrm{d}r} -T- V
\end{split}
\end{equation}

where we have assumed that the total mass is simply the volume integral of the density profile

\begin{equation}
 M = \int_\mathrm{0}^R {4 \pi r^2 \rho(r) \mathrm{d}r }.
\end{equation}

Solving Eq. (\ref{rest}) for $T+V$, we obtain

\begin{equation}
 T+V = \int_\mathrm{0}^R {4 \pi r^2 e^b \epsilon(r) \mathrm{d}r} + \int_\mathrm{0}^R {4 \pi r^2 \rho(r) \left( 1- e^b \right) \mathrm{d}r}.
\end{equation}

This leads to the definition

\begin{eqnarray}
 T &=& \int_\mathrm{0}^R {4 \pi r^2 \epsilon(r) e^b \mathrm{d}r} \\
 U &=& \int_\mathrm{0}^R {4 \pi r^2 \rho(r) \left(1-e^b\right) \mathrm{d}r}.
\end{eqnarray}

Consider a top-hat density profile and a two component fluid consisting of DM and a cosmological constant such that

\begin{equation}
 e^{2b} = \frac{1}{1-Ar^2} \qquad \textnormal{with} \qquad A= \frac{8 \pi G}{3} \left(\rho+ \rho_\Lambda \right).
\end{equation}

Thus, we finally end up with:

\begin{eqnarray}
  T &=& \int_\mathrm{0}^R {4 \pi r^2 \epsilon \frac{1}{\sqrt{1-Ar^2}}\mathrm{d}r} \\
  U &=& \int_\mathrm{0}^R {4 \pi r^2 \rho \left(1- \frac{1}{\sqrt{1-Ar^2}} \right) \mathrm{d}r}.
\end{eqnarray}

\section{Static, spherically-symmetric field equations with homogeneous DE}\label{app_C}

Consider a two component fluid described by

\begin{eqnarray}
 T_{\mu\nu}^{(m)}&=& \left(\rho+p\right) u_{\mu} u_{\nu} + p g_{\mu\nu} \\
 T_{\mu\nu}^{(Q)}&=& \left(\rho_Q+p_Q\right) u_{\mu} u_{\nu} + p_Q g_{\mu\nu} \\ \nonumber
\\
T_{\mu\nu} &=& T_{\mu\nu}^{(m)} + T_{\mu\nu}^{(Q)}
\end{eqnarray}

where the densities $\rho$ and $\rho_Q$ are assumed to be constant and the quintessence component has an equation of state $p_Q=w\rho_Q$ with constant $w$. Energy-momentum conservation is fulfilled separately for each fluid component:

\begin{eqnarray}
 \nabla_{\mu}T^{\mu\nu, m}&=&0 \\
 \nabla_{\mu}T^{\mu\nu, Q}&=&0.
\end{eqnarray}

The static, spherically symmetric field equations of this set-up are

\begin{eqnarray} \nonumber
 G_{\mu\nu}=8\pi G T_{\mu\nu}\\ \nonumber \\
\label{w_Feqs 1}
\frac{1}{r^2}-e^{-2b}\left(\frac{1}{r^2}-\frac{2b'}{r}\right) &=& 8\pi G \left(\rho + \rho_Q \right) \\
\label{w_Feqs 2}
-\frac{1}{r^2}+e^{-2b}\left(\frac{1}{r^2}+\frac{2a'}{r}\right) &=& 8\pi G \left( p + w\rho_Q \right)\\
\label{w_Feqs 3}
e^{-2b}\left(a''-a'b'+a'^2+\frac{a'-b'}{r}\right) &=& 8\pi G \left( p+ w \rho_Q \right).
\end{eqnarray}

 We have to find out whether there are conditions for the solvability of the field equations without using any concrete solution for $a$ and $b$ such that effects resulting from boundary conditions are excluded.\\

With the help of Eq. (\ref{w_Feqs 1}) we can express $b'$: 

\begin{equation}
\label{b'}
 b' = \frac{1}{2r} \left(1 - \left(1- 8 \pi G \left(\rho +\rho_Q\right) r^2\right) e^{2b} \right).
\end{equation}

In the same way Eq. (\ref{w_Feqs 2}) can be solved for $a'$:

\begin{equation}
\label{a'}
 a' = -\frac{1}{2r} \left(1- \left(1+ 8 \pi G \left(p + w\rho_Q\right) r \right)e^{2b}\right).
\end{equation}

If we add Eq. (\ref{w_Feqs 1}) and Eq. (\ref{w_Feqs 2}), we will get 

\begin{equation}
 a' = -b' + 4 \pi G \left( \rho + p +\rho_Q (1+w)\right)r e^{2b}.
\end{equation}

If Eq. (\ref{b'}) is inserted into that expression, we will obtain Eq. (\ref{a'}) so the first two field equations are consistent.\\

Energy-momentum conservation for the matter component of the fluid means

\begin{equation}
 \nabla_\mu T^{\mu\nu, m} =0.
\end{equation}

Projecting this onto the space perpendicular to the velocity flow, we obtain the relativistic Euler equation

\begin{equation}
 \left(g_{ \alpha\nu} + u_{\alpha}u_{\nu}\right)\nabla_\mu T^{\mu\nu} =0 
\end{equation}

which becomes

\begin{equation}
 \left(\rho+p\right) \nabla_u u = -\textnormal{grad}p -u \nabla_u p.
\end{equation}

In case of a static configuration, we are left with 

\begin{equation}
\label{hydr}
 - p' = a'\left(\rho+p\right)
\end{equation}

which is basically the hydrostatic equilibrium condition for the matter configuration of our system.\\

Consider the derivative of Eq. (\ref{a'})

\begin{equation}
\label{a''0}
\begin{split}
   a'' = &\frac{1}{2r^2} \left(1+ e^{2b}\left\{-1 + 8 \pi G \left(p + w\rho_Q\right)r^2 + 8 \pi G p' r^3  + 2b'r \right. \right.\\ 
         & \left. \left. \left(1+ 8\pi G \left(p+w\rho_Q\right)\right)r^2 \right\} \right).
\end{split}
\end{equation}

Inserting Eq. (\ref{b'}), Eq. (\ref{a'}) and Eq. (\ref{hydr}) into Eq. (\ref{a''0}) and simplifying leads to

\begin{equation}
\label{a''1}
\begin{split}
  a'' = &\frac{1}{2r^2}+\frac{1}{2r^2}e^{2b} 4 \pi G \left(5p+4 w \rho_Q +\rho\right) r^2 \\
        &-\frac{1}{2r^2} e^{4b} \left(1+8\pi G \left(p + w\rho_Q\right)r^2\right)\\
        \ &\left(1- 4\pi G \left(\rho-p+2\rho_Q\right)r^2\right).
\end{split}
\end{equation}

On the other hand $a''$ can be expressed with the help of Eq. (\ref{w_Feqs 3}) 

\begin{equation}
 a''= a'b' - a'^2 - \frac{a'-b'}{r} + 8\pi G \left(p + w\rho_Q \right) e^{2b}.
\end{equation}

If we plug in Eqs. (\ref{b'}) and (\ref{a'}), we can obtain (after some algebra)

\begin{equation}
\label{a''2}
 \begin{split}
  a'' = &\frac{1}{2r^2}+\frac{1}{2r^2}e^{2b} \left[ 4 \pi G \left(\rho+ \rho_Q\right)r^2+20 \pi G \left(p + w\rho_Q\right)r^2\right]\\
       -&\frac{1}{2r^2}e^{4b}\left(1+8\pi G \left(p+ w \rho_Q\right)r^2\right) \left(1- 4 \pi G \left(\rho + \rho_Q \right. \right.\\
       -& \left. \left. p - w\rho_Q\right)\right).
 \end{split}
\end{equation}

For Eqs. (\ref{w_Feqs 1}) - (\ref{w_Feqs 3}) being consistent, Eq. (\ref{a''1}) and Eq. (\ref{a''2}) have to be identical which means that each coefficient belonging to the same order of $e^{2b}$ has to be equal for each radius $r$. \\

\begin{setlength}{\leftmargini}{0.3cm}
\begin{itemize}
 \item 0th order:
\begin{equation*} 
 \frac{1}{2r^2}= \frac{1}{2r^2} \qquad \textnormal{trivially fulfilled}
\end{equation*}
\item 1st order:
\begin{equation*}
\begin{split} 
  &\frac{1}{2r^2} 4 \pi G \left(5p + 4w\rho_Q +\rho \right) r^2 \\ 
= &\frac{1}{2r^2} \left[ 4 \pi G \left(\rho + \rho_Q \right) r^2 + 20 \pi G \left(p + w\rho_Q \right) r^2\right] 
\end{split}
\end{equation*}
\begin{eqnarray*} 
 \Rightarrow 5p + \rho + 4w \rho_Q &=& \rho + \rho_Q + 5p + 5w\rho_Q \\ 
 \Rightarrow 4w  &=&  \left( 1+5w \right) \\ 
 \Rightarrow w&=&-1
\end{eqnarray*}
\item 2nd order:
\begin{equation*} 
\begin{split}
 &\frac{1}{2r^2} \left(1+ 8 \pi G \left(p +w \rho_Q\right)r^2\right)\left(1- 4 \pi G \left(\rho -p +2\rho_Q\right)r^2\right) \\
=&\frac{1}{2r^2} \left(1+ 8 \pi G \left(p +w \rho_Q\right)r^2\right)\left(1- 4 \pi G \left(\rho -p +\rho_Q -w \rho_Q\right)r^2\right)
\end{split}
\end{equation*}
If we require that $p \neq -w\rho_Q -1/(8\pi G r^2)$ which we have to assume in the general case, we can say
\begin{eqnarray*} 
 4\pi G \left(\rho -p +2\rho_Q\right) &=& 4 \pi G \left(\rho - p +\rho_Q -w\rho_Q\right)\\ 
\Rightarrow 2 \rho_Q &=& \rho_Q \left(1-w\right) \\  
\Rightarrow w&=&-1 
\end{eqnarray*}

\end{itemize}
\end{setlength}

Thus,  Eq. (\ref{w_Feqs 1}) - Eq. (\ref{w_Feqs 3}) necessarily require $w = -1$ in order to be consistently solvable.

\section{Iterative method to find the virial redshift}\label{app_D}

Consider Eq. (\ref{nlm}) and solve this for the non-linear density contrast $\delta(a)$. Once this is done, Eq. (\ref{Delta_V}) can be used

\begin{equation}
 \Delta(r, a) \equiv \frac{\rho(r)}{\rho_{\mathrm{b}}(a)} = \zeta \left(\frac{x}{y} \right)^3 = 1+ \delta(a).
\end{equation}

This allows to express $y(a)$ with the help of $\delta$

\begin{equation}
\label{y2}
 y(a)= \frac{a}{a_{ \mathrm{ta}}} \cdot \left(\frac{\zeta}{1+\delta(a)}\right)^{1/3}.
\end{equation}

Using Eq. (\ref{y2}) and the virialization equation (Eqs. (\ref{vir_eq}), (\ref{virialization_eq}), (\ref{virialization_PN})), an iterative method can be constructed to find the virial scale factor $a_{\mathrm{vir}}$. 

Starting from $a^{(0)}=a_{\mathrm{c}}$, we will proceed in the following way

\begin{equation} 
 \begin{split}
 &a^{(0)} \overset{(\ref{vir_eq})} {\longrightarrow} y_{\mathrm{vir}}^{(0)} \overset{(\ref{y2})} {\longrightarrow} a\left(y_\mathrm{vir}^{(0)}\right)=a^{(1)}
\\ 
 &a^{(1)} \overset{(\ref{vir_eq})} {\longrightarrow} y_{\mathrm{vir}}^{(1)} \overset{(\ref{y2})} {\longrightarrow} a\left(y_\mathrm{vir}^{(1)}\right)=a^{(2)}  \\ 
 &a^{(2)} \overset{(\ref{vir_eq})} {\longrightarrow} y_{\mathrm{vir}}^{(2)} \overset{(\ref{y2})} {\longrightarrow} a\left(y_\mathrm{vir}^{(2)}\right)=a^{(3)} \\ 
 & \qquad \qquad  \vdots \qquad \qquad \vdots \qquad  \qquad \vdots \\
 &a^{(n)} \overset{(\ref{vir_eq})} {\longrightarrow} y_\mathrm{vir}^{(n)} \overset{(\ref{y2})} {\longrightarrow} a\left(y_\mathrm{vir}^{(n)}\right)=a^{(n+1)}.  \\ 
 \end{split}
\end{equation}

In each step, the quantity $a(y^{(i)})$ is needed which is given by the root of the equation

\begin{equation}
\label{root}
 y(a)-y_\mathrm{vir}^{(i)}=0 \longrightarrow a(y_\mathrm{vir}^{(i)})
\end{equation}

and which can be found numerically in each step $i$.\\

It turns out that this method converges\footnote{Given a tolerance of $tol \sim 10^{-7}$, the iteration can be stopped within 3 to 4 steps.}, so that the following condition can be applied to stop the iteration: 

\begin{equation}
\label{condition}
 \left|\frac{a^{(n+1)}-a^{(n)}}{a^{(n)}}\right| \leq tol
\end{equation}

with an appropriate tolerance $tol$. Thus, the final result can be defined as:

\begin{equation}
 a^{(n+1)}\equiv a_{\mathrm{vir}} \Rightarrow z_{\mathrm{vir}} = \frac{1}{a_{\mathrm{vir}}}-1.
\end{equation}\\

It will turn out that the virial overdensity changes significantly if the exact quantities are evaluated at $z=z_\mathrm{vir}$:

Let us calculate $\Delta_V$ at redshift $z_{\mathrm{c}}=0$ in the EdS model which corresponds to a virial redshift of $z_{\mathrm{vir}}=0.065$ and a turn-around redshift of $z_{\mathrm{ta}}=0.587$.\\

\begin{setlength}{\leftmargini}{0.3cm}
\begin{itemize}
 \item Case 1 ($z=z_\mathrm{c}$): \\
  \begin{equation}
  \label{case1}
   \Delta_V=\left(\frac{3 \pi}{4}\right)^2 \left(1+z_{\mathrm{ta}}\right)^3 \left(\frac{1}{y_{\mathrm{vir}}}\right)^3= 177.518
  \end{equation}
 \item Case 2 ($z=z_\mathrm{vir}$):  \\
  \begin{equation}
  \label{case2}
   \Delta_V=\left(\frac{3 \pi}{4}\right)^2 \left(\frac{1+z_{\mathrm{ta}}}{1+z_{\mathrm{vir}}}\right)^3 \left(\frac{1}{y_{\mathrm{vir}}}\right)^3= 146.958
  \end{equation}
\end{itemize}
\end{setlength}

This has also been predicted analytically by Lee, Ng (2010) (see \cite{lee_spherical_2010}).

\end{document}